\documentclass[11pt,letterpaper]{article}
\usepackage[nomarkers,nolists,figuresonly]{endfloat}

\usepackage[latin1]{inputenc}

\usepackage{color}
\usepackage{graphics}
\usepackage{graphicx}
\usepackage{times}
\usepackage[T1]{fontenc}
\usepackage{amsmath}
\usepackage{amssymb}
\usepackage{amsthm}
\usepackage{tcolorbox}
\usepackage{ae,aecompl}
\usepackage{textcomp}   % for degree (\textdegree) symbol
\usepackage{hyperref}
\hypersetup{colorlinks=true}
\usepackage{caption}    % for figure
\usepackage{subcaption} %  and subfigures

\usepackage{transparent}

\usepackage{fancybox}

\usepackage{tabularx}
\newcolumntype{L}[1]{>{\hsize=#1\hsize\raggedright\arraybackslash}X}%
\newcolumntype{R}[1]{>{\hsize=#1\hsize\raggedleft\arraybackslash}X}%
\newcolumntype{C}[1]{>{\hsize=#1\hsize\centering\arraybackslash}X}%

\newcommand{\clrr}[1]{\textcolor{red}{#1}}

\newcommand{\clrlg}[1]{\textcolor{green!100!black!95}{#1}}
\newcommand{\clrb}[1]{\textcolor{blue}{#1}}

\newcommand{\clrm}[1]{\textcolor{magenta}{#1}}
\newcommand{\clrp}[1]{\textcolor{violet}{#1}}
\newcommand{\clrc}[1]{\textcolor{cyan}{#1}}

\def\grapprox{\mathrel{\rlap{\lower3.5pt\hbox{$\mathchar"218$}}\raise 
1pt \hbox{$\mathchar"13E$}}}
\def\lsapprox{\mathrel{\rlap{\lower3.5pt\hbox{$\mathchar"218$}}\raise 
1pt \hbox{$\mathchar"13C$}}}

\definecolor{clr1}{RGB}{0,255,255}
\definecolor{clr3}{RGB}{255,0,255}

\begin{document}

\title{
\vspace{1.75in}
\textbf{ITER-IA 3D MHD Simulations of \\Shattered Pellet Injection(SPI)}\\
\vspace{0.1in}
\textbf{D1.1} Optimization of the SPI model \\
\vspace{0.2in}
\normalsize{
in partial fulfillment of ITER Agreement Ref: IO/IA/20/4300002130\\
to the Agreement on Scientific Cooperation Ref: LGA-2019-A-73}
\vspace{0.1in}
}

%\author{C.~C.~Kim - SLS2 Consulting
%}
\author{Charlson.~C.~Kim\\
SLS2 Consulting\\ San Diego USA\\
Email: kimcc@fusion.gat.com\\[0.05in]
B.~C.~Lyons, Y.~Q.~Liu, J.~T.~McClenaghan,\\ P.~B.~Parks, L.~L.~Lao\\
General Atomics\\ San Diego USA\\[0.05in]
%V.~A.~Izzo\\
%Fiat Lux\\ San Diego USA\\[0.05in]
%R.~W.~Harvey, Y.~V.~Petrov\\
%CompX\\ Oceanside USA\\[0.05in]
%M.~Lehnen, A.~Loarte\\
%ITER\\ Saint-Paul-lez-Durance France\\[0.075in]
%\& the NIMROD Team
}
\date{}

\clearpage
\maketitle
\thispagestyle{empty}
\newpage

\graphicspath{
	{./Figures/}
%	{../ITERDMS/031920meeting/}
%	{../ITER030921/}
%	{../ITER040621/}
%	{../CTTS0620/}
	}

\section*{Axisymmetric SPI Simulation Scenarios 1-5}

This report is in partial fulfillment of deliverable \textbf{D1.1 Optimization of the SPI model} and summarizes
axisymmetric ITER SPI parameter scans performed by the NIMROD code for several ITER equilibria.  3D scans of the
toroidal extent will be included in subsequent reports to be submitted within the next few months.

These axisymmetric parameter scans are to assess the sensitivity of various injection parameters in preparation for 3D
MHD SPI simulations.  The scans are comprised of 5 scenarios:
\begin{itemize}
	\item[\bf S1]- fragment size scan : 3 uniform pencil beam, 1 distributed size pencil beam (table~\ref{tab:shatparam})
\begin{table}[h]
	\begin{subtable}{0.5\textwidth}
\centerline{
	\begin{tabular}{|l|c|c|c|}
	\hline
	r$_{frag}$(mm) & \clrp{0.5}   & \clrc{2.5} & \clrlg{5.0} \\ \hline
	S              & \clrp{40}    & \clrc{8}   & \clrlg{ 4 } \\ \hline
	N              & \clrp{64000} & \clrc{512} & \clrlg{64 } \\ \hline
	N$_{marker}$   & \clrp{800}   & \clrc{32}  & \clrlg{32 } \\ \hline
	n$_{actual}$   & \clrp{80}    & \clrc{16}  & \clrlg{ 2 } \\ \hline
	\end{tabular} 
}
	\caption{uniform pencil beam composition}
	\label{tab:unipb}
	\end{subtable}
	\begin{subtable}{0.5\textwidth}
\centerline{
	\begin{tabular}{|l|c|c|c|c|}
        \hline
		r$_{frag}$(mm) &\clrr{ 0.5} &\clrr{ 2.5} &\clrr{ 5.0} &\clrr{ 7.5 }\\ \hline
		$M_f/M_p$(\%)  &\clrr{0.39} &\clrr{21.1} &\clrr{46.9} &\clrr{ 31.6}\\ \hline
		N              &\clrr{ 250} &\clrr{ 108} &\clrr{ 30 } &\clrr{  6  }\\ \hline
		N$_{marker}$   &\clrr{  50} &\clrr{  27} &\clrr{ 30 } &\clrr{  6  }\\ \hline
		n$_{actual}$   &\clrr{  5 } &\clrr{   4} &\clrr{  1 } &\clrr{  1  }\\ \hline
        \end{tabular}
}
	\caption{distributed size pencil beam composition}
	\label{tab:dispb}
	\end{subtable}
	\caption{Plume composition for {\bf S1} Fragment Size Scan: the three uniform pencil beam plumes
	(\ref{tab:unipb}) listing fragment radii (r$_{frag}$), shatter parameter (S=r$_{pellet}$/r$_{frag}$), number of
	fragments (N), number of simulation marker particles (N$_{marker}$), and number of fragments each marker
	particle represents (n$_{actual}$). N=n$_{actual}\times$N$_{marker}$. Table~\ref{tab:dispb} shows the
	composition of the distributed size pencil beam plume, approximating a realistic shatter, listing the fragment
	radii (r$_{frag}$) and mass fraction ($M_f/M_p$(\%)) in lieu of the shatter parameter.}
	\label{tab:shatparam}
\end{table}
	\item[\bf S2]- velocity scan : v = [\clrb{250},\clrlg{500},\clrr{750}]m/s
	\item[\bf S3]- velocity dispersion scan : dv/v = [\clrb{0.2},\clrr{0.4}] (linear distribution)
	\item[\bf S4]- poloidal extent of plume : d$\theta_{hw}$ = [\clrb{15$^{\circ}$},\clrr{45$^{\circ}$}] (linear distribution) (dv/v=0.2)
	\item[\bf S5]- poloidal injection angle : $\theta$ = $\pm$[\clrc{2}\clrb{0}$^{\circ}$,\clrm{4}\clrr{5}$^{\circ}$] (dv/v=0.2)
\end{itemize}

These scans are performed with several ITER equilibria representative of the operating range, from low current and
thermal energy (H123 5MA, 29MJ Hydrogen H-mode) to high current and high thermal energy (DT24 15MA, 370MJ D-T H-mode).
The four equilibria are summarized in table~\ref{tab:ITERrefscan}.  A more detailed description can be found here on the
ITER document server (IDM): \href{https://user.iter.org/?uid=4UKXDJ}{ITER\_D\_4UKXDJ}.  These equilibria were originally
generated by S.~H.~Kim (sunhee.kim@iter.org) and subsequently improved to higher resolution and reduced error by
J.~T.~McClenaghan (mcclenaghanj@fusion.gat.com).  The original equilibria were used for scenarios 1,2, while the
improved equilibria were used for scenarios 3-5.  For the latter scenarios 3-5, the H123 H-mode was not considered due
to numeric issues related to its unique equilibrium profiles.  The short scale length structure near the core of the
current profile proved particularly troublesome in producing the high accuracy equilibrium reconstructions necessary for
NIMROD simulations.  Residual Grad-Shafranov errors produced unphysical numeric forces that resulted in singular-like
behavior.  Higher resolutions may have remedied this but required more computing time than available for these survey
scans.

\begin{table}[h]
\centerline{
	\begin{tabular}{|l|c|c|c|c|}
	\hline
	         &   DT24 D-T H-mode   &  H26 Hydrogen L-mode  &  He56 He H-mode  &  H123 Hydrogen H-mode \\ \hline
	equil.@  &   400s          &    60s       &   400s        &   100s       \\ \hline
	Itot     &   15MA          &   15MA       &   7.5MA       &   5MA        \\ \hline
	TE       &   370MJ         &   35MJ       &   57MJ        &   29MJ       \\ \hline
	\end{tabular}
}
	\caption{ITER equilibria : time of reconstruction, total current, thermal energy.}
	\label{tab:ITERrefscan}
\end{table}

For all scenarios, the nominal pellet is 5\% Neon+95\% Deuterium, diameter D=28.5mm, with length to diameter ratio of L/D=2.
A spherical pellet of r$_{pellet}$=20.0mm is equivalent to this cylindrical pellet.  All fragments are computed with
respect to the equivalent spherical pellet.  For scenarios 3-5, the DT24 equilibria use a spherical pellet of r$_{pellet}$=25.0mm
to minimize burn through.

Unless otherwise specified, the plume is composed of identical fragments with radius r$_{frag}$=2.5mm and velocity
v=500m/s and a poloidal injection angle of 0$^{\circ}$.  Fragment radius r$_{frag}$=2.5mm with a shatter parameter ($\equiv$pellet
radius/fragment radius) S=8 results in N$_{fragments}$=512 fragments, except for DT24
Scenarios 3-5, where S=10 and N$_{fragments}$=1000.

\begin{figure}
\centerline{
\includegraphics[width=1.0\textwidth,trim={18.0cm 3.00cm 15.0cm 2.0cm},clip]{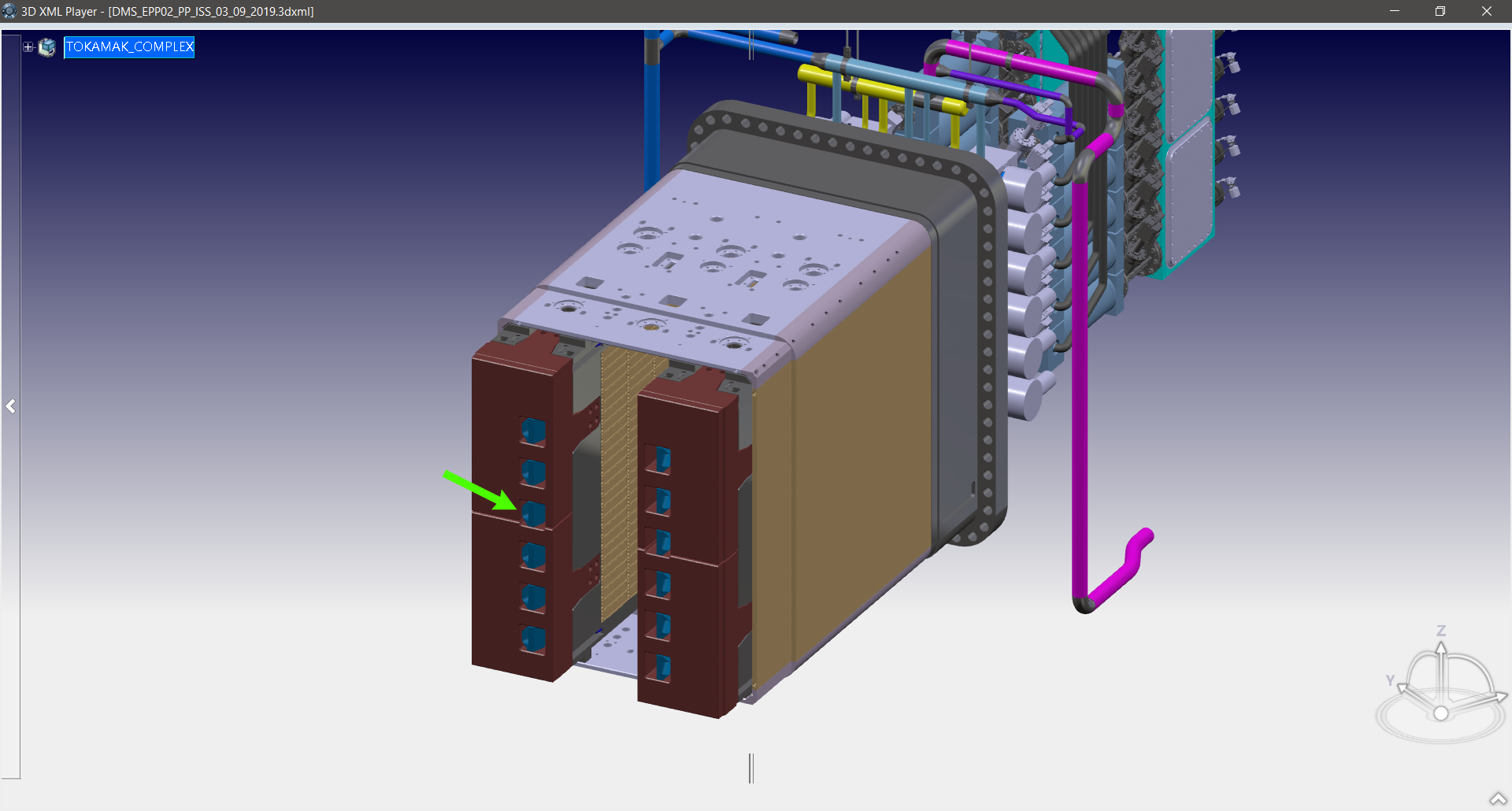}
}
	\caption{ITER Equatorial Shattered Pellet Injector showing two columns of 6 injectors.  The nominal injector
	used in the simulation is EQ\_08\_4\_1, fourth from the bottom, left column.  This typically aligns a bit above
	the magnetic axis.}
	\label{fig:spiinjector}
\end{figure}

Figure~\ref{fig:spiinjector} shows a rendering of the ITER Equatorial Shattered Pellet Injector.  All fragments are
injected from equatorial port EQ\_08\_4\_1 located at R=8.454233m and Z=0.6855m.  This injector typically lies several
centimeters above the magnetic axis.  

High density diffusion is used for stability but may sometimes have consequences for energy conservation :
1.0-4.0m$^2$/s for Scenarios 1,2, 10.0m$^2$/s for Scenarios 3-5.  Viscosity values used were 2000-5000m$^2$/s for
Scenarios 1,2, 10000m$^2$/s for Scenarios 3-5.  A Spitzer-like temperature (T$^{-3/2}$) dependent resistivity model is
used with a minimum floor value.  The minimum electric diffusivity (=resistivity/$\mu_0$) 0.1m$^2$/s for Scenarios 1,2
and 1.0-2.0m$^2$/s for Scenarios 3-5.

The thermal quench time at the q=2 surface (q=2 $\tau_{TQ}$) is determined by the time at which T$_{q=2}$=10eV for
Scenario 1.  For scenarios 3-5, T$_{q=2}$=10eV for equilibria H26, T$_{q=2}$=20eV for He56, and T$_{q=2}$=50eV for DT24.
The differing temperatures reflect the need to set the open flux/boundary temperatures to different values depending on
the thermal energy content of the equilibria to mitigate negative temperatures that can sometimes occur in the
simulations.  For scenarios 1,2 all open flux/boundary temperatures are 5eV.  For scenarios 3-5, open flux/boundary
temperatures are 5eV for H26, 10eV for He56, and 20eV for DT24.  The global thermal quench time is read directly from the thermal
energy plots.

Radiation fractions listed in the tables are not quite accurate.  All simulations include both Ohmic and viscous
heating.  For Scenario 1, the radiation fraction is reported at t=2.0ms and includes all radiation.  Scenarios 2-5
report the radiation at the thermal quench time accounting for Ohmic contributions.  However, due to an oversight, the
viscous heating was not kept track of.  In lieu of the viscous heating contribution to the radiation, the kinetic energy
is used in its place, resulting in not quite accurate radiation fractions.  However, the trends should still hold.  For
most cases, the viscous heating is small.  We expect viscous heating to be smaller than the kinetic energy.  The kinetic
energy, in turn, is smaller than the magnetic energy by about a factor of $\beta$.  We expect Ohmic heating dominates viscous
heating in most cases.

\section{S1 - Fragment Size Scan : r$_{pb}$=[\clrlg{5.0},\clrc{2.5},\clrp{0.5}]mm, \clrr{mixed distribution}}

\begin{table}
\centerline{
\begin{tabular}{|l|c|c|c|c|} \hline
	        scenario 1                &  DT24 H-mode  &  H26 L-mode  &  He56 H-mode  &  H123 H-mode  \\ \hline
                thermal energy   	  &     370MJ     &     35MJ     &     57MJ      &     29MJ      \\ \hline \hline
             rf@2ms \clrr{md}    	  &      0.18     &     0.23     &     0.24      &     0.66      \\ \hline
            rf@2ms \clrlg{5.0mm} 	  &      0.18     &     0.21     &     0.21      &     0.59      \\ \hline
             rf@2ms \clrc{2.5mm} 	  &      0.25     &     0.37     &     0.42      &     1.28      \\ \hline
             rf@2ms \clrp{0.5mm} 	  &      0.32     &     0.61     &     0.51      &     ***       \\ \hline \hline
          assim@2ms \clrr{md}    	  &      0.46     &     0.08     &     0.13      &     0.08      \\ \hline
          assim@2ms \clrlg{5.0mm}	  &      0.45     &     0.06     &     0.12      &     0.08      \\ \hline
          assim@2ms \clrc{2.5mm} 	  &      0.61     &     0.09     &     0.19      &     0.13      \\ \hline
          assim@2ms \clrp{0.5mm} 	  &      0.91     &     0.20     &     0.30      &     ***       \\ \hline \hline
	q=2 $\tau_{TQ}$(ms) \clrr{md}     &      3.7      &     1.3      &     2.2       &     2.0       \\ \hline
	q=2 $\tau_{TQ}$(ms) \clrlg{5.0mm} &      3.8      &     1.4      &     2.6       &     2.1       \\ \hline
	q=2 $\tau_{TQ}$(ms) \clrc{2.5mm}  &      2.3      &     1.1      &     1.5       &     1.3       \\ \hline
	q=2 $\tau_{TQ}$(ms) \clrp{0.5mm}  &      1.4      &     ***      &     0.8       &     0.8       \\ \hline
\end{tabular}}
\caption{Thermal quench metrics for {\bf S1} Fragment Size listing radiation fraction and assimilation fraction at
	t=2.0ms and thermal quench time at the q=2 surface for each ITER scenario.  ``***'' denotes unavailable data.}
	\label{tab:tqmetrics}
\end{table}

\begin{figure}
\centerline{
	\hspace{0.00cm}
	\begin{subfigure}[b]{0.50\textwidth}
		\includegraphics[width=1.0\textwidth]{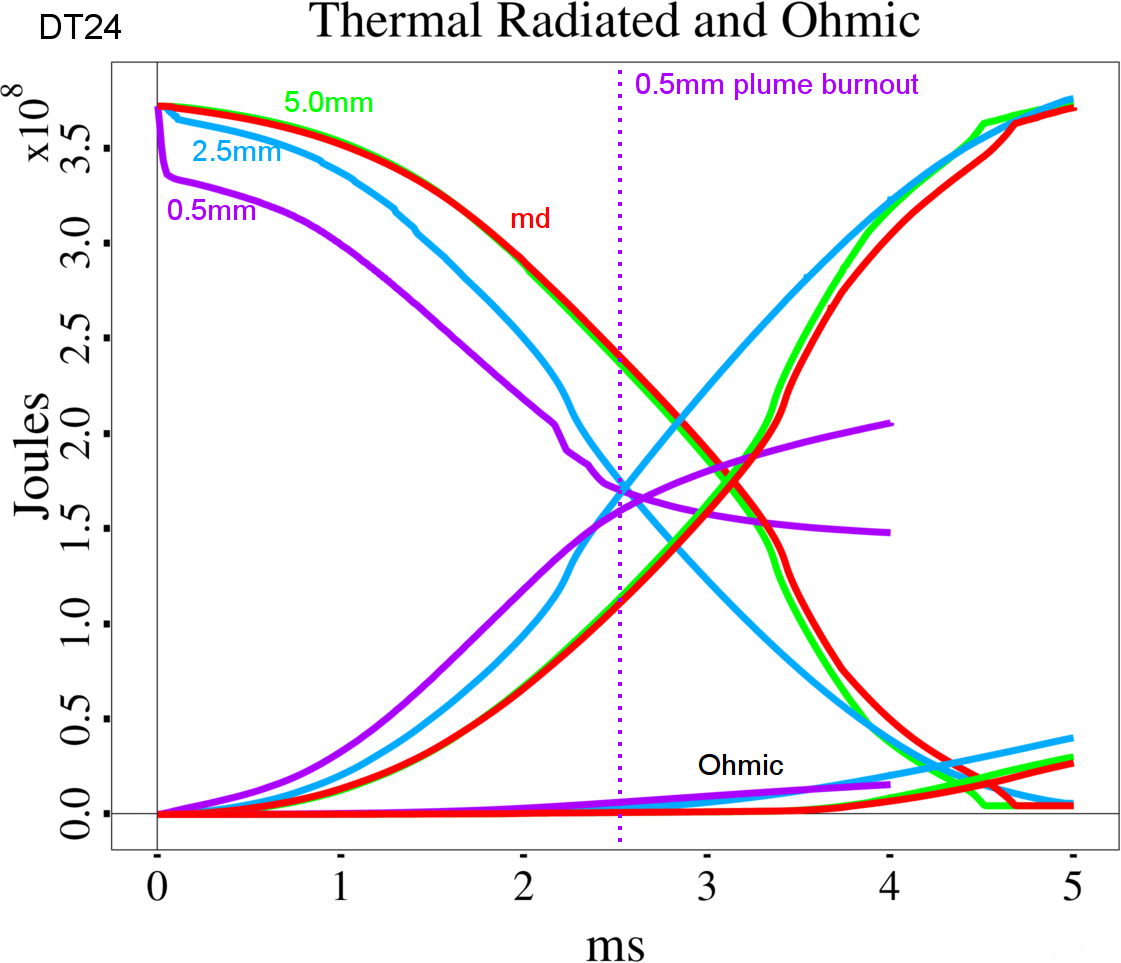}
		\caption{DT24 (15MA)}
		\label{fig:TEDT24_rsc}
	\end{subfigure}
	\hspace{0.00cm}
	\begin{subfigure}[b]{0.50\textwidth}
		\includegraphics[width=1.0\textwidth]{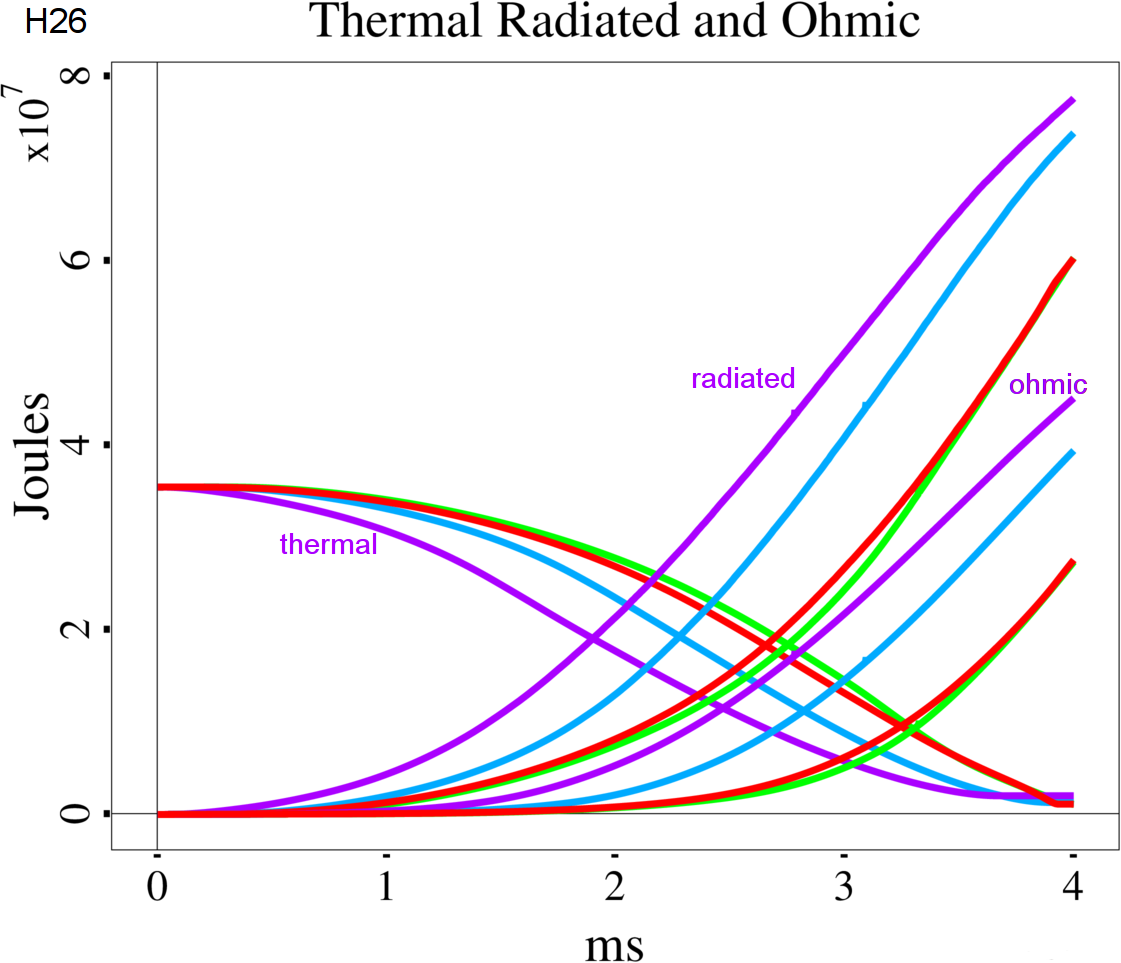}
		\caption{H26 L-mode (15MA)}
		\label{fig:TEH26l_rsc}
	\end{subfigure}
}
	\vspace{0.25cm}
\centerline{
	\hspace{0.00cm}
	\begin{subfigure}[b]{0.50\textwidth}
		\includegraphics[width=1.0\textwidth]{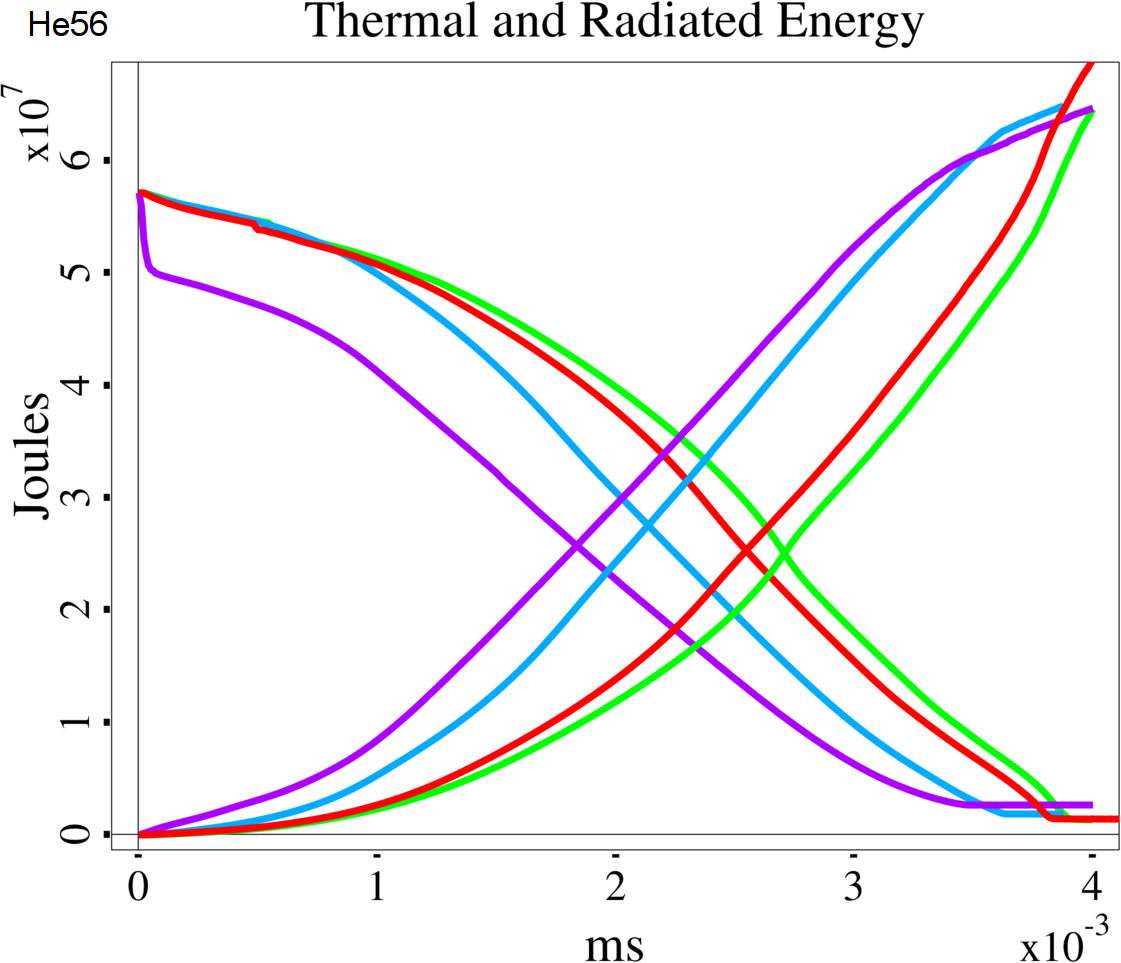}
		\caption{He56}
		\label{fig:TEHe56_rsc}
	\end{subfigure}
	\hspace{0.00cm}
	\begin{subfigure}[b]{0.50\textwidth}
		\includegraphics[width=1.0\textwidth]{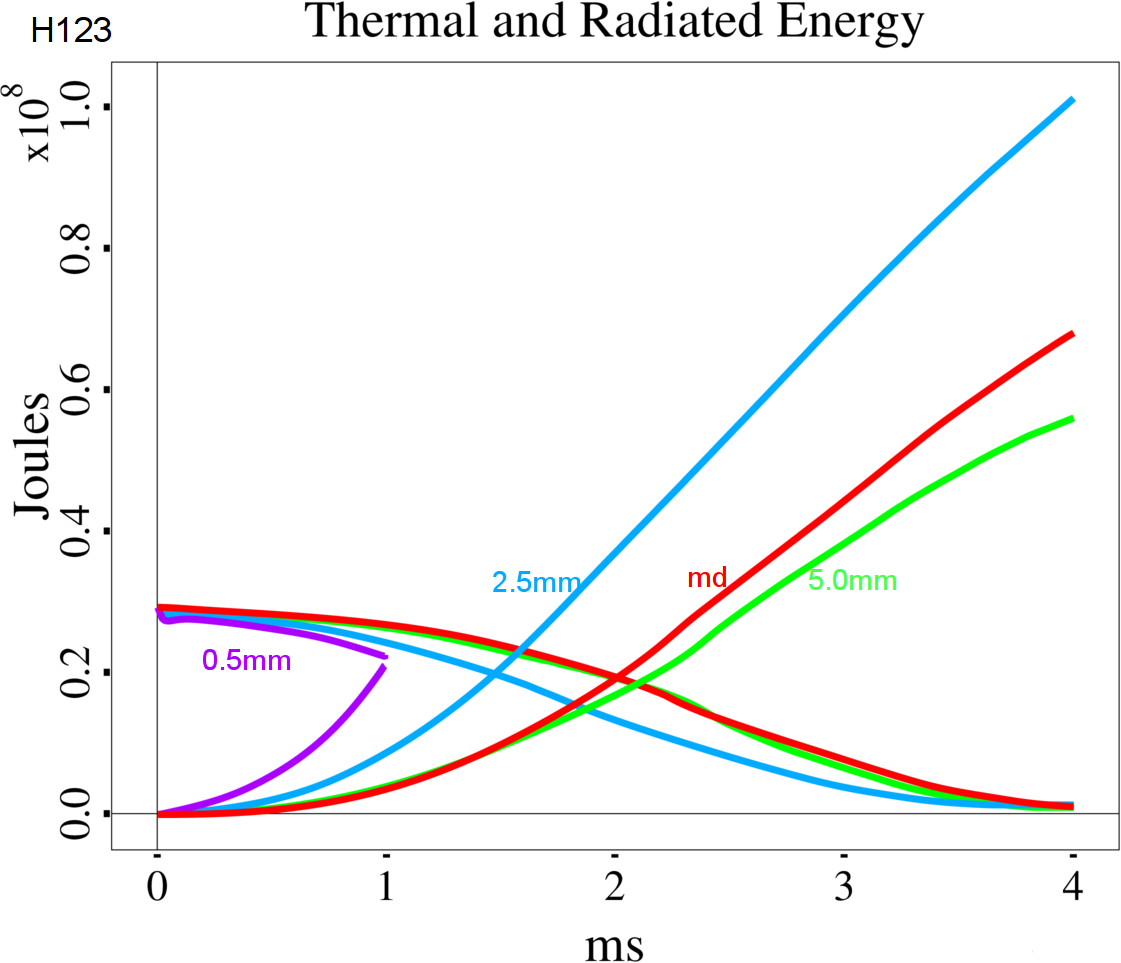}
		\caption{H123}
		\label{fig:TEH123_rsc}
	\end{subfigure}
	}
	\caption{Comparison of {\bf S1} Fragment Size : thermal and radiated energies for the three uniform pencil
	beams and distributed pencil beam show faster thermal quench and more radiation for smaller fragment size plumes.
	(\ref{fig:TEDT24_rsc}) and (\ref{fig:TEH26l_rsc}) also plot the Ohmic energy to contrast relative contributions
	for the high and low thermal energy case. (\ref{fig:TEDT24_rsc}) shows plume burn through for r=0.5mm.
	(\ref{fig:TEH123_rsc}) shows numeric termination for r=0.5mm.}
	\label{fig:TE_rsc}
\end{figure}

For all four {\bf S1} cases considered, the SPI simulation particles\cite{kimc:2019} are arranged in a pencil beam distribution; particles
uniformly distributed along a line co-linear with the trajectory with a length of 40.0cm.  This is equivalent to 0.80ms
at the plume velocity of v=500m/s.  For the mixed distribution, each set of markers for a particular radius is uniformly
distributed along the 40.0cm.  The mixed distribution approximates the Park's statistical shatter model\cite{parks2016}.

The fragment size scan shows that the smaller fragment size plume produces more ablation
resulting in more radiation.  Although individually, a fragment's ablation rate decreases with size, collectively their
number scales faster than the decreasing ablation rate.  There are 1000 times more fragments in a \clrp{r=0.5mm, S=40}
plume than in a \clrlg{r=5.0mm, S=4} plume (table~\ref{tab:unipb}).  The high ablation rate for \clrp{r=0.5mm, S=40}
causes the early dip in thermal energy seen for each of the H-mode equilibria; a result of the pedestal collapsing.

Figures~\ref{fig:TE_rsc} show a comparison of thermal and radiated energies as the fragment size is
varied for each equilibria.  Plots for DT24(\ref{fig:TEDT24_rsc}) and H26(\ref{fig:TEH26l_rsc}) also include the Ohmic
energy to contrast the relative contributions for the high and low thermal energy cases.  The Ohmic thermal energy
contributes to the overall radiation.  Consequently, the late radiation has a larger Ohmic component for the lower
thermal energy content equilibria H26(\ref{fig:TEH26l_rsc}) and H123(\ref{fig:TEH123_rsc}).

DT24(\ref{fig:TEDT24_rsc}) \clrp{r=0.5mm, S=40} shows complete burn through of the plume at t=2.5ms resulting in an
incomplete thermal quench.  H123(\ref{fig:TEH123_rsc}) \clrp{r=0.5mm, S=40} terminated early due to numeric
instabilities.

For all equilibria, the \clrr{mixed distribution} pencil beam most closely tracks the evolution of the uniform
\clrlg{r=5.0mm, S=4}. This is not too surprising since almost 50\% of the mass is in r=5.0mm fragments
(table~\ref{tab:dispb}), more than twice that of r=2.5mm fragments and orders of magnitude more than r=0.5mm fragments.
Although the r=7.5mm fragments hold $\sim$30\% of the mass, this mass is constrained to only 6 fragments; as explained
above, not enough ablators.

Table~\ref{tab:tqmetrics} lists the radiation and assimilation fractions at t=2.0ms and the thermal quench time for the
q=2 surface. This table shows that the smaller fragment radius plume ablates more and causes more radiation and a faster
thermal quench.

\begin{figure}
\centerline{
	\hspace{0.00cm}
	\begin{subfigure}[b]{0.275\textwidth}
		\includegraphics[width=1.0\textwidth]{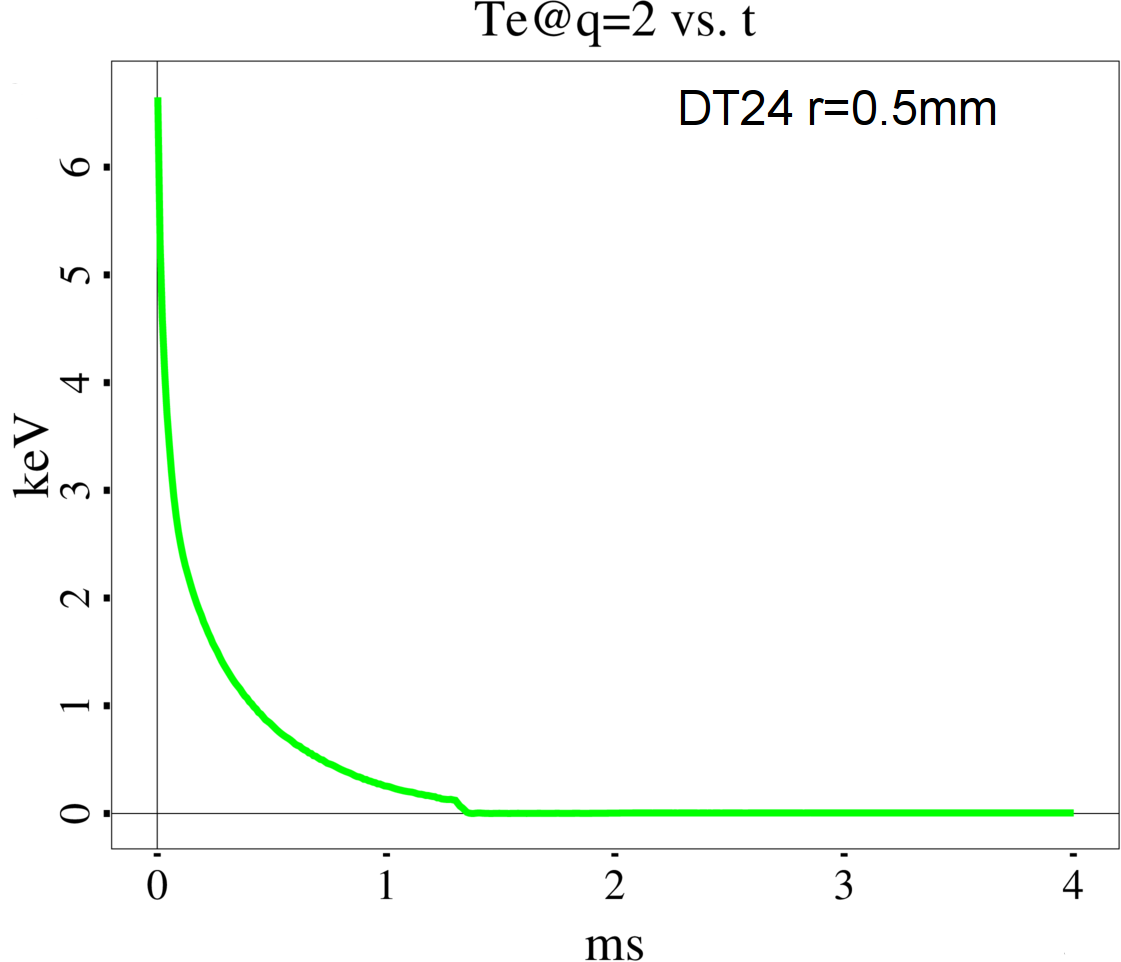}
		\caption{DT24 r=0.5mm}
		\label{fig:DT24_r05}
	\end{subfigure}
	\hspace{0.00cm}
	\begin{subfigure}[b]{0.275\textwidth}
		\includegraphics[width=1.0\textwidth]{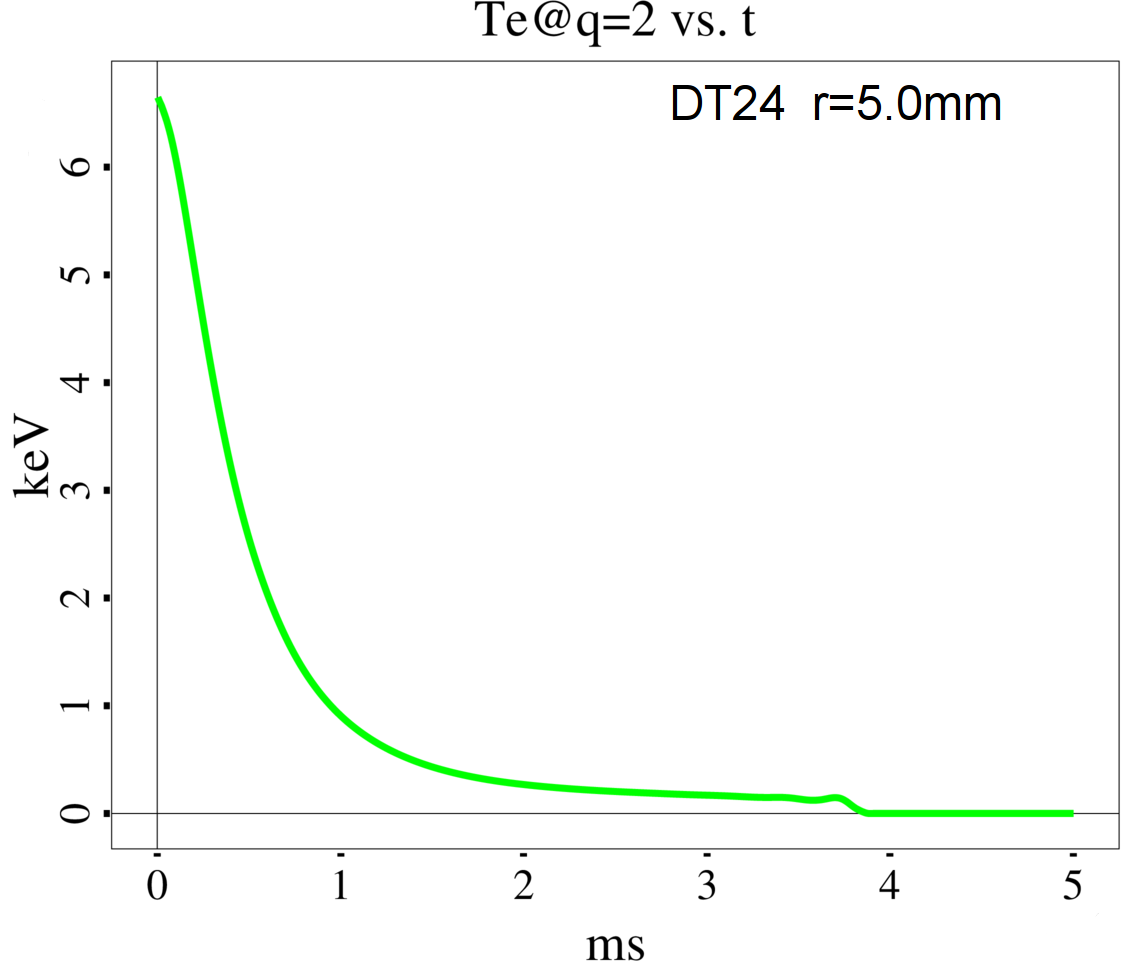}
		\caption{DT24 r=5.0mm}
		\label{fig:DT24_r50}
	\end{subfigure}
	\hspace{0.00cm}
	\begin{subfigure}[b]{0.275\textwidth}
		\includegraphics[width=1.0\textwidth]{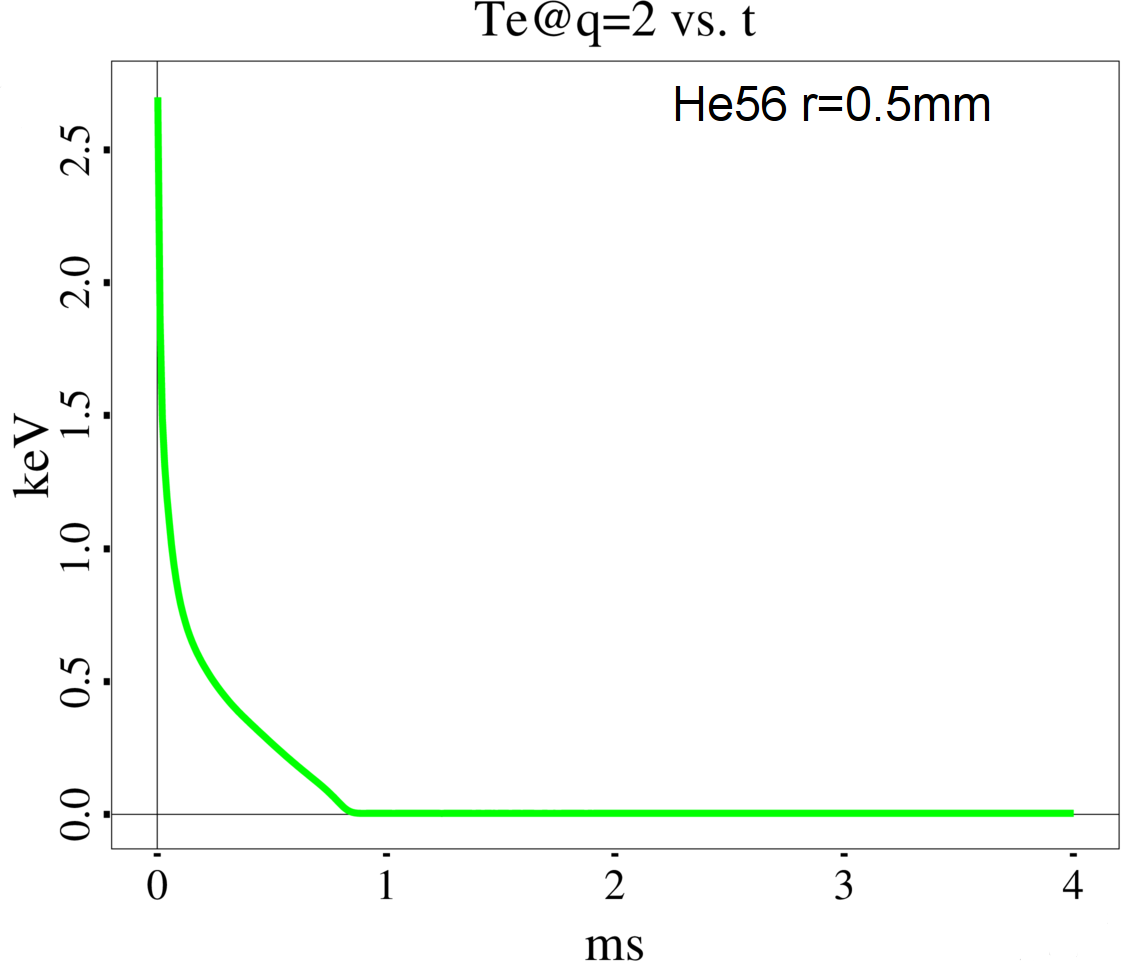}
		\caption{He56 r=0.5mm}
		\label{fig:He56_r05}
	\end{subfigure}
	\hspace{0.00cm}
	\begin{subfigure}[b]{0.275\textwidth}
		\includegraphics[width=1.0\textwidth]{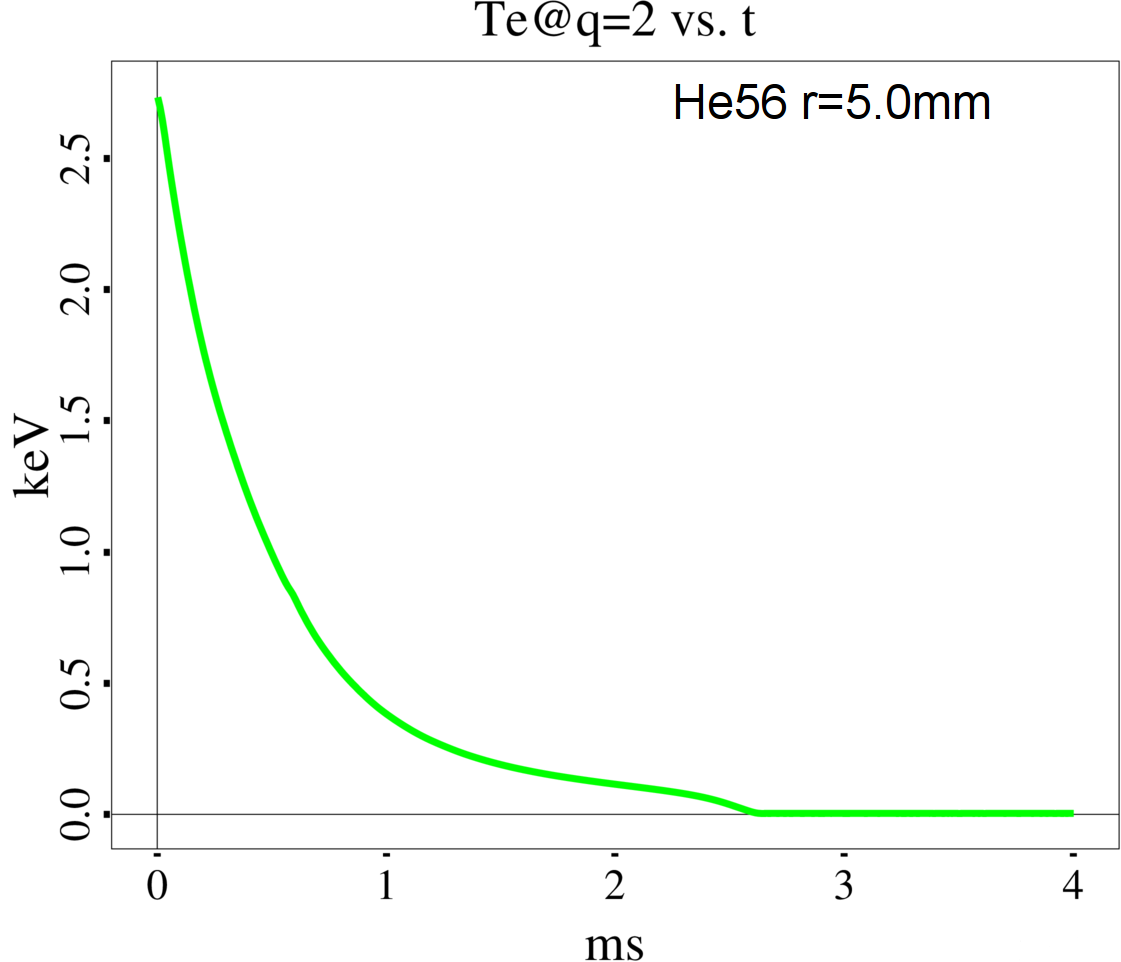}
		\caption{He56 r=5.0mm}
		\label{fig:He56_r50}
	\end{subfigure}
	} 
	\caption{Temperature at the q=2 surface.  Note the longer tails (\ref{fig:DT24_r50},\ref{fig:He56_r50}) for
	larger fragment radius plumes caused by their lower ablation rates.}
	\label{fig:tq2}
\end{figure}

Figure~\ref{fig:tq2} shows a comparison of q=2 thermal quench time for DT24 and He56 for r$_f$=[0.5,5.0]mm.  These plots
show that the longer q=2 thermal quench times for the mixed distribution and \clrlg{5.0mm} fragment plume are due to
lingering tails in the temperature.  The large fragment plumes have a low ablation rate and require more time to
completely quench, indicating a lower quench efficiency.  It is anticipated that the smaller fragment plumes
will be more perturbative in the 3D simulations, increasing the likelihood of triggering a large MHD event.

\section{S2 - Plume Velocity Scan : v=[\clrb{250},\clrlg{500},\clrr{750}]m/s, r$_f$=2.5mm}

\begin{figure}
\centerline{
	\hspace{0.00cm}
	\begin{subfigure}[b]{0.375\textwidth}
		\includegraphics[width=1.0\textwidth]{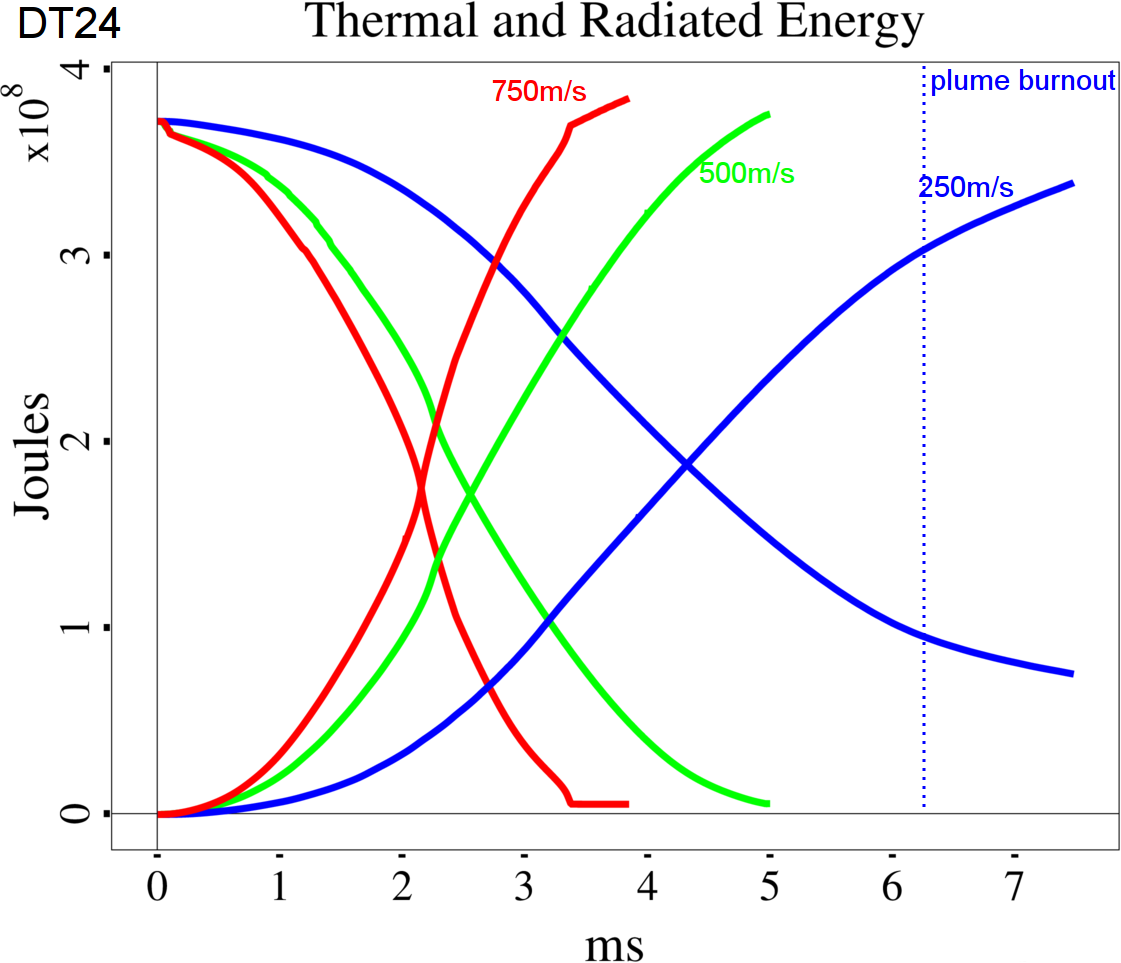}
		\caption{DT24}
		\label{fig:DT24_vscan}
	\end{subfigure}
	\hspace{0.00cm}
	\begin{subfigure}[b]{0.375\textwidth}
		\includegraphics[width=1.0\textwidth]{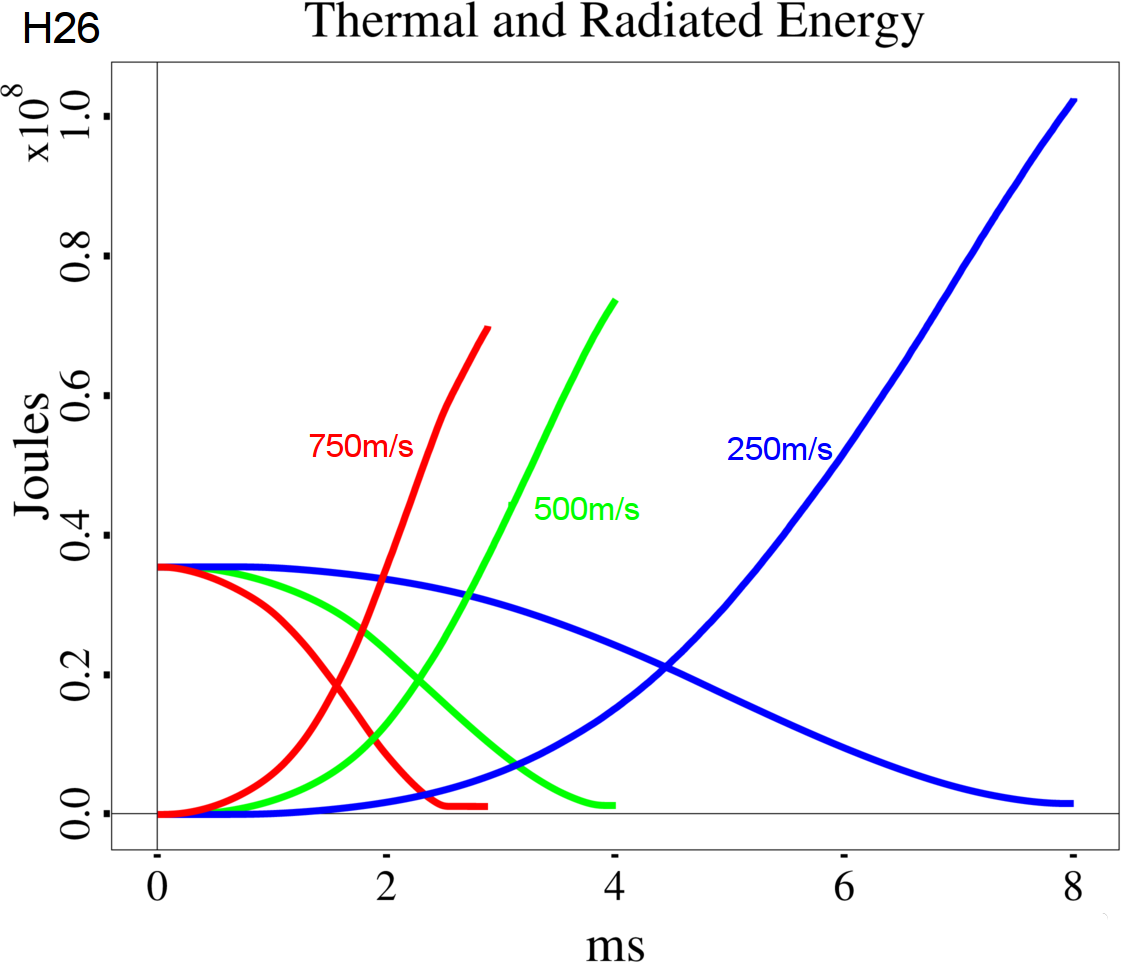}
		\caption{H26 L-mode}
		\label{fig:H26l_vscan}
	\end{subfigure}
%}
%	\vspace{0.25cm}
%\centerline{
	\hspace{0.00cm}
	\begin{subfigure}[b]{0.375\textwidth}
		\includegraphics[width=1.0\textwidth]{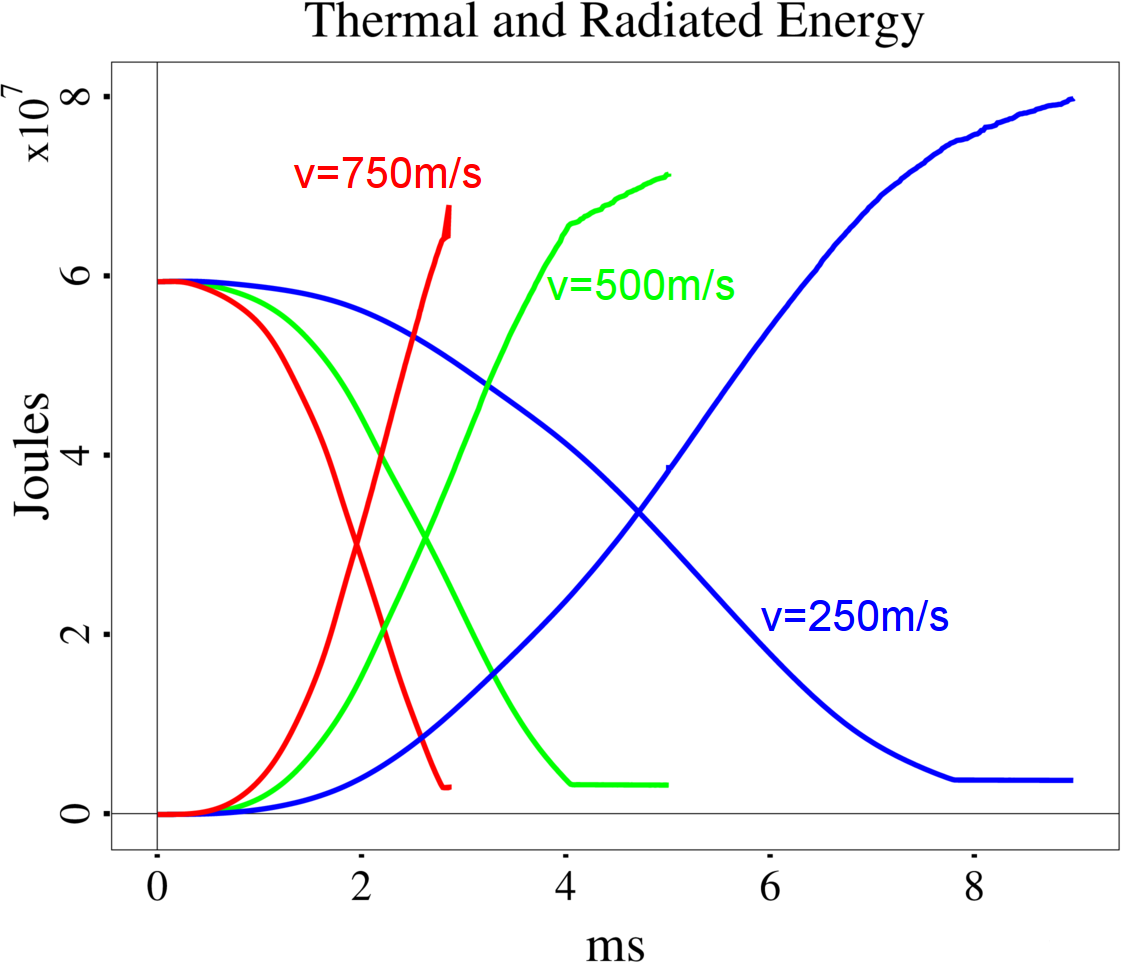}
		\caption{He56}
		\label{fig:He56_vscan}
	\end{subfigure}
%	\hspace{0.00cm}
%	\begin{subfigure}[b]{0.50\textwidth}
%		\includegraphics[width=1.0\textwidth]{RadPowHe56}
%		\caption{He56}
%		\label{fig:He56_vscanRP}
%	\end{subfigure}
	}
%	\caption{Thermal and radiated energy for {\bf S2} Velocity shows $\tau_{TQ}\propto 1/v_{plume}$. Note the `knee's
%	in He56(\ref{fig:He56_vscan}) for the two sets of v=250m/s curves. Increasing density diffusion delays the knee.}
	\caption{Thermal and radiated energy for {\bf S2} Velocity shows $\tau_{TQ}\propto 1/v_{plume}$. DT24 \clrb{v=250m/s} shows plume burn through at t=6.24ms.}
	\label{fig:vscan}
\end{figure}

The pencil beams for each velocity maintain the same time separation between each simulation marker particle
resulting in shorter pencil beams for slower plumes; (\clrb{20.0},\clrlg{40.0},\clrr{60.0})cm for
(\clrb{250},\clrlg{500},\clrr{750})m/s.

Figure~\ref{fig:vscan} shows faster plumes quench faster and thermal quench time is approximately inversely proportional
to the plume velocity.  It can be inferred from the radiation energy traces that the faster plumes radiate more
intensely.

It is anticipated that in 3D simulations, faster fragments will be more perturbative.
\section{S3 - Velocity Dispersion Scan : dv/v=[\clrb{0.2},\clrr{0.4}]($\equiv$[\clrb{$\pm$100},\clrr{$\pm$200}]m/s)}

\begin{table}
	\begin{subtable}{1.0\textwidth}
\centerline{
\begin{tabular}{|r||c|c|c|}\hline
	H26 scenario 3 		& \clrlg{p.b.}     & \clrb{dv/v=0.2} & \clrr{dv/v=0.4} \\ \hline \hline
	$\tau_{TQ}$(ms)		&    	4.24       &       3.76      &       3.44      \\ \hline
	assim.			&    	0.223      &       0.218     &       0.189     \\ \hline
	rad. frac.		&    	0.97       &       0.97      &       0.96      \\ \hline
	q=2 $\tau_{TQ}$(ms)	&       1.49       &       1.18      &       1.15      \\ \hline
	q=2 assim.              &       0.043      &       0.023     &       0.035     \\ \hline
\end{tabular}
}
	\caption{H26 L-mode}
	\label{tab:s3H26TQM}
\end{subtable}

	\vspace{0.25cm}
	\begin{subtable}{1.0\textwidth}
\centerline{
\begin{tabular}{|r||c|c|c|}\hline
	He56 scenario 3		&      \clrlg{p.b.} & \clrb{dv/v=0.2} & \clrr{dv/v=0.4} \\ \hline \hline
	$\tau_{TQ}$(ms)		&       4.04        &        3.56     &       3.34      \\ \hline
	assim.			&       0.334       &        0.335    &       0.296     \\ \hline
	rad. frac.		&       0.90        &        0.86     &       0.89      \\ \hline
	q=2 $\tau_{TQ}$(ms)     &       1.64        &        1.25     &       1.26      \\ \hline
	q=2 assim.              &       0.111       &        0.109    &       0.112     \\ \hline
\end{tabular}
}
	\caption{He56}
	\label{tab:s3He56TQM}
\end{subtable}

	\vspace{0.25cm}
	\begin{subtable}{1.0\textwidth}
\centerline{
\begin{tabular}{|r||c|c|c|}\hline
	DT24 scenario 3		&      \clrlg{p.b.} & \clrb{dv/v=0.2} & \clrr{dv/v=0.4} \\ \hline \hline
	$\tau_{TQ}$(ms)		&      4.30         &       3.90      &      4.04       \\ \hline
	assim.			&      0.718        &       0.726     &      0.651      \\ \hline
	rad. frac.		&      0.87         &       0.85      &      0.87       \\ \hline
	q=2 $\tau_{TQ}$(ms)	&      1.87         &       1.44      &      1.48       \\ \hline
	q=2 assim.         	&      0.367        &       0.375     &      0.388      \\ \hline
\end{tabular}
}
	\caption{DT24}
	\label{tab:s3DT24TQM}
\end{subtable}
	\caption{Thermal Quench Metrics for {\bf S3} Velocity Dispersion listing thermal quench time($\tau_{TQ}$),
	assimilation and radiation fractions at t=$\tau_{TQ}$ and q=2 quench time and assimilation show faster thermal quench when
	assimilation<0.5 but slower when assimilation>0.5.}
	\label{tab:s3TQM}
\end{table}

\begin{figure}
\centerline{
	\hspace{0.00cm}
	\begin{subfigure}[b]{0.375\textwidth}
		\includegraphics[width=1.0\textwidth]{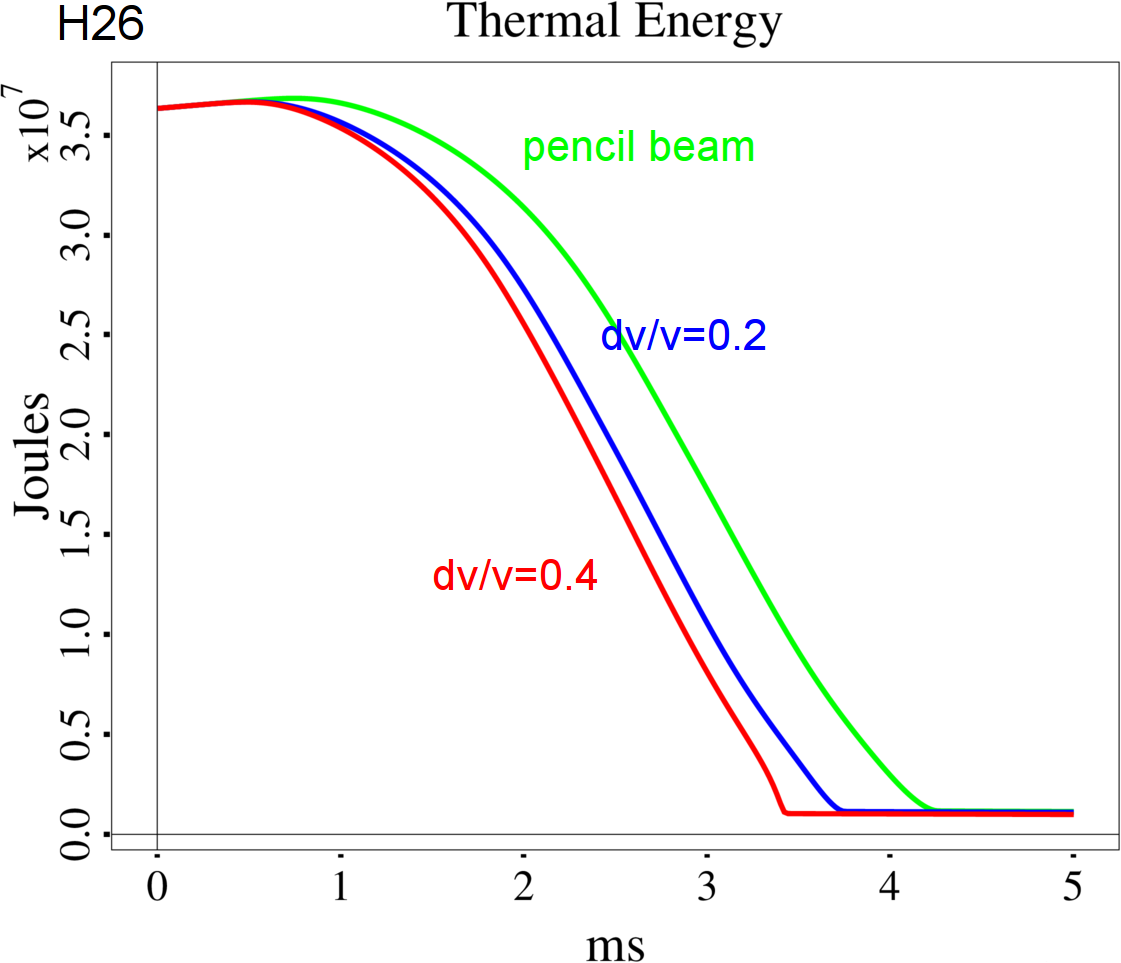}
		\caption{H26 L-mode}
		\label{fig:H26_TES3}
	\end{subfigure}
	\hspace{0.00cm}
	\begin{subfigure}[b]{0.375\textwidth}
		\includegraphics[width=1.0\textwidth]{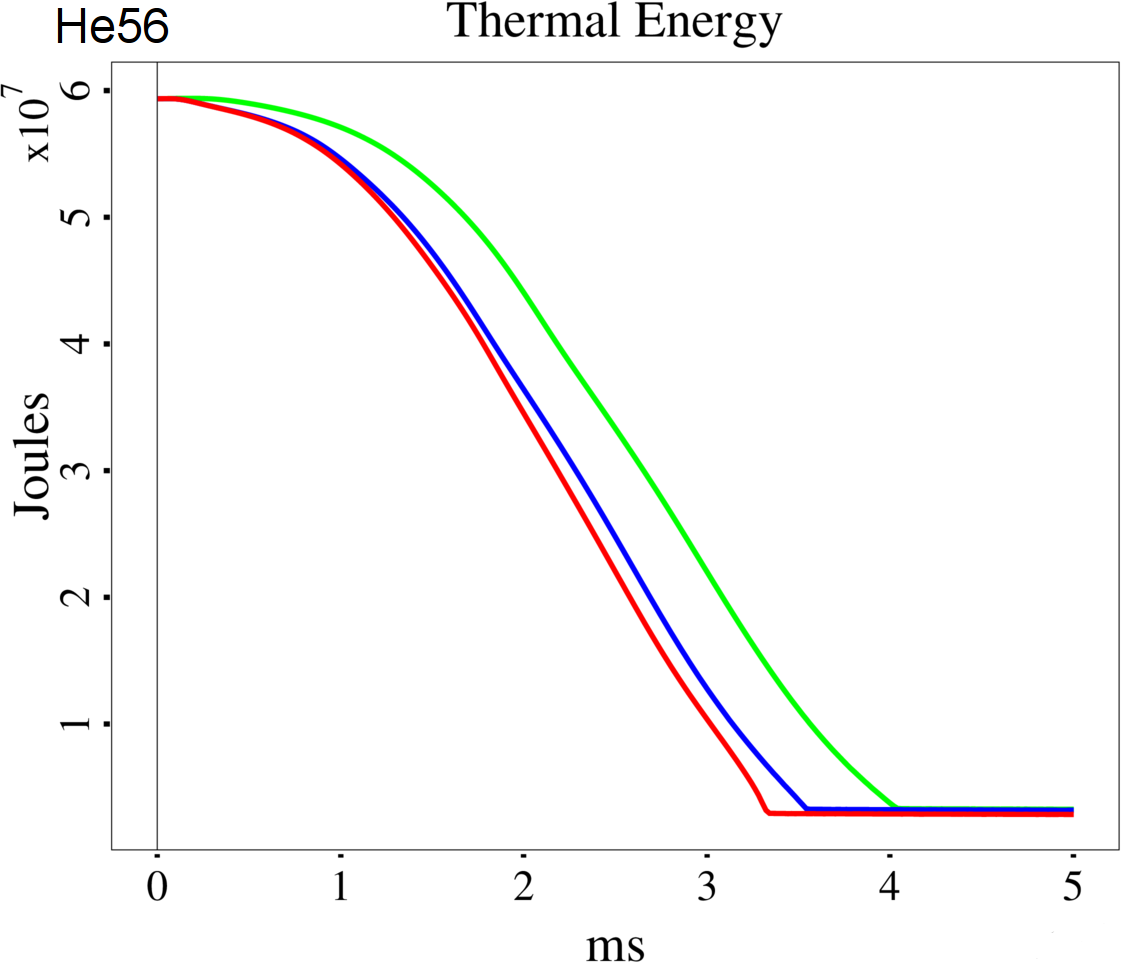}
		\caption{He56}
		\label{fig:He56_TES3}
	\end{subfigure}
	\hspace{0.00cm}
	\begin{subfigure}[b]{0.375\textwidth}
		\includegraphics[width=1.0\textwidth]{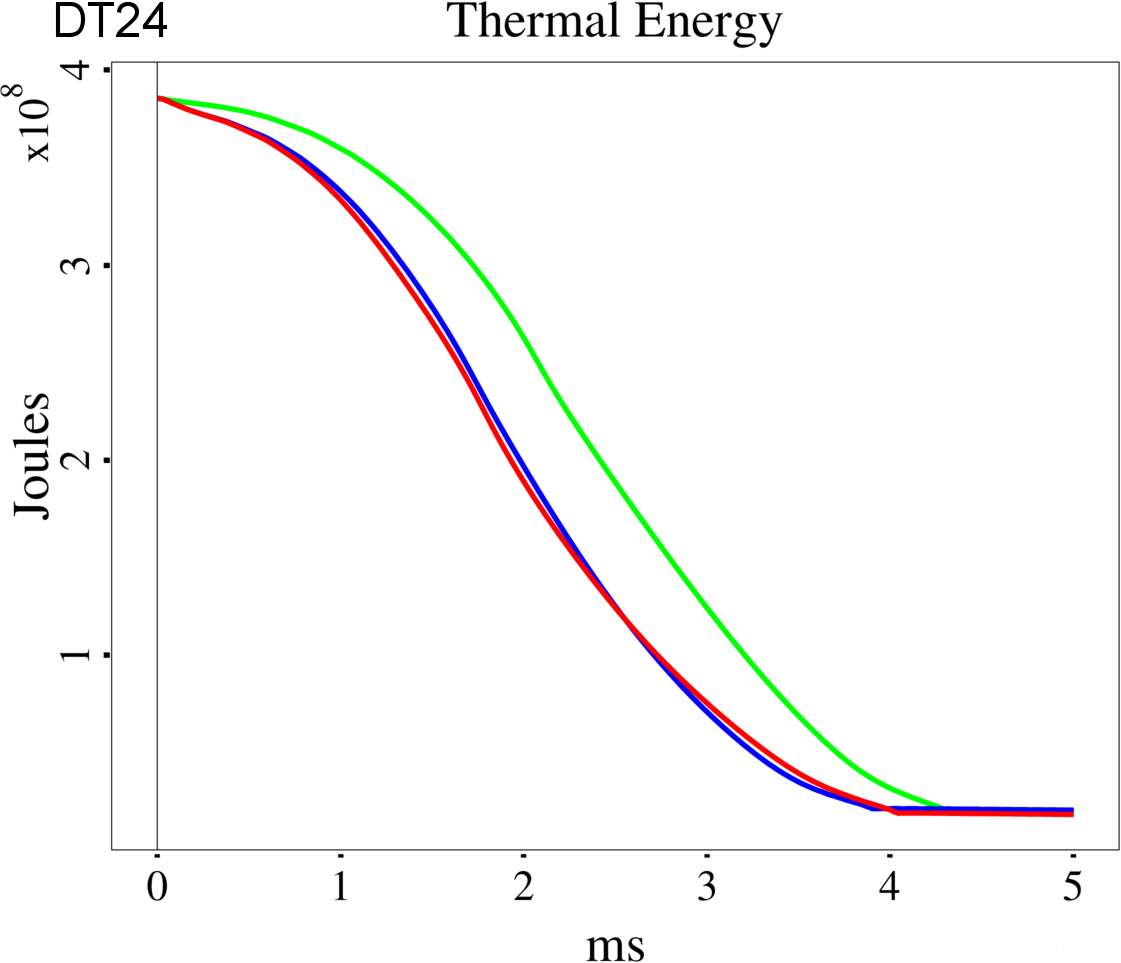}
		\caption{DT24}
		\label{fig:DT24_TES3}
	\end{subfigure}
}
	\caption{Comparison of thermal energy for {\bf S3} Velocity Dispersion shows faster thermal quench consistent with
	faster fragments.  DT24 (\ref{fig:DT24_TES3}) shows cross-over of the velocity dispersion curves at t=2.55ms.  Due
	to the high assimilation fraction, the slower back half of \clrr{dv=0.4} is overtaken by \clrb{dv=0.2}.}
	\label{fig:TES3}
\end{figure}
The remaining three scenarios simulate all 2.5mm radius fragments; 512 markers for the S=8 cases H26,He54 and 1000 for
S=10 DT24.  For the velocity dispersion scan the trajectories are co-linear.  Fragment velocities  are randomly assigned
from a uniform distribution between v=[v$_0$-dv,v$_0$+dv].  Although a Gaussian distribution may be more representative
of a physical distribution, we choose the uniform distribution to better assess the impact of the velocity dispersion
because it equally weights the full distribution.  The r=2.5mm, 40.0cm pencil beam plume is included as a reference.

Figures~\ref{fig:TES3} and tables~\ref{tab:s3TQM} show that when assimilation is less than 0.50, the higher dispersion
quenches faster, whereas, when assimilation is greater than 0.50, the lower dispersion quenches faster.  This is
consistent with the observation from Scenario 2 that faster fragments quench faster.  The higher dispersion plume has
fastest fragments in the forward half but also the slowest fragments in the back half.  DT24 figure~\ref{fig:DT24_TES3}
shows a cross-over of the two velocity dispersion thermal energy curves at t=2.55ms as the slower back half of
dv=0.4 is overtaken.

The higher velocity dispersion results in lower assimilation; perhaps due to the broader extent of the fragment beam.

Comparison with the pencil beam plume indicates that the 40.0cm beam is too long.  A 20.0cm pencil beam may be more equivalent.

It is anticipated that in 3D simulations, lower velocity dispersion, narrower plumes will be more perturbative.

For the remaining two scenarios, the plumes will have a velocity v=500m/s and a velocity dispersion dv/v=0.2.

\section{S4 - Poloidal Extent : d$\theta_{hw}$=[\clrb{15$^{\circ}$},\clrr{45$^{\circ}$}] (dv/v=0.2)}

\begin{table}
	\begin{subtable}{1.0\textwidth}
\centerline{
\begin{tabular}{|r||c|c|c|}\hline
	H26 scenario 4		&\clrlg{d$\theta$=0$^{\circ}$}&\clrb{d$\theta$=15$^{\circ}$}&\clrr{d$\theta$=45$^{\circ}$}\\ \hline \hline
	$\tau_{TQ}$(ms)		&           3.76              &        4.00                 &          4.58               \\ \hline
	assim.			&           0.218             &        0.219                &          0.200              \\ \hline
	rad. frac.		&           0.97              &        0.96                 &          0.96               \\ \hline
	q=2 $\tau_{TQ}$(ms)	&           1.18              &        1.19                 &          1.28               \\ \hline
	q=2 assim.         	&           0.023             &        0.034                &          0.034              \\ \hline
\end{tabular}
	}
	\caption{H26 L-mode}
	\label{tab:s4H26ltqm}
\end{subtable}

	\vspace{0.25cm}
	\begin{subtable}{1.0\textwidth}
\centerline{
\begin{tabular}{|r||c|c|c|}\hline
	He56 scenario 4		&\clrlg{d$\theta$=0$^{\circ}$}&\clrb{d$\theta$=15$^{\circ}$}&\clrr{d$\theta$=45$^{\circ}$}\\ \hline \hline
	$\tau_{TQ}$(ms)		&           3.56              &          3.84               &          4.62               \\ \hline
	assim.			&           0.335             &          0.336              &          0.314              \\ \hline
	rad. frac.		&           0.86              &          0.92               &          0.98               \\ \hline
	q=2 $\tau_{TQ}$(ms)	&           1.25              &          1.26               &          1.37               \\ \hline
	q=2 assim.         	&           0.109             &          0.108              &          0.110              \\ \hline
\end{tabular}
	}
	\caption{He56}
	\label{tab:s4He56tqm}
\end{subtable}

	\vspace{0.25cm}
	\begin{subtable}{1.0\textwidth}
\centerline{
\begin{tabular}{|r||c|c|c|}\hline
	DT24 scenario 4		&\clrlg{d$\theta$=0$^{\circ}$}&\clrb{d$\theta$=15$^{\circ}$}&\clrr{d$\theta$=45$^{\circ}$}\\ \hline \hline
	$\tau_{TQ}$(ms)		&            3.90             &           4.16              &          5.00*              \\ \hline
	assim.			&            0.726            &           0.727             &          0.665              \\ \hline
	rad. frac.		&            0.85             &           0.87              &          0.90               \\ \hline
	q=2 $\tau_{TQ}$(ms)	&            1.44             &           1.45              &          1.56               \\ \hline
	q=2 assim.              &            0.375            &           0.373             &          0.364              \\ \hline
\end{tabular}
	}
	\caption{DT24}
	\label{tab:s4DT24tqm}
\end{subtable}
	\caption{Thermal Quench Metrics for {\bf S4} Poloidal Extent listing thermal quench time($\tau_{TQ}$),
	assimilation and radiation fractions at t=$\tau_{TQ}$ and q=2 quench time and assimilation shows increasing thermal quench time with increasing
	angle due to decrease (by $\sim\cos{\Delta\theta}$) of effective normal velocity component of the plume.}
	\label{tab:s4tqm}
\end{table}

\begin{figure}
\centerline{
	\hspace{0.00cm}
	\begin{subfigure}[b]{0.375\textwidth}
		\includegraphics[width=1.0\textwidth]{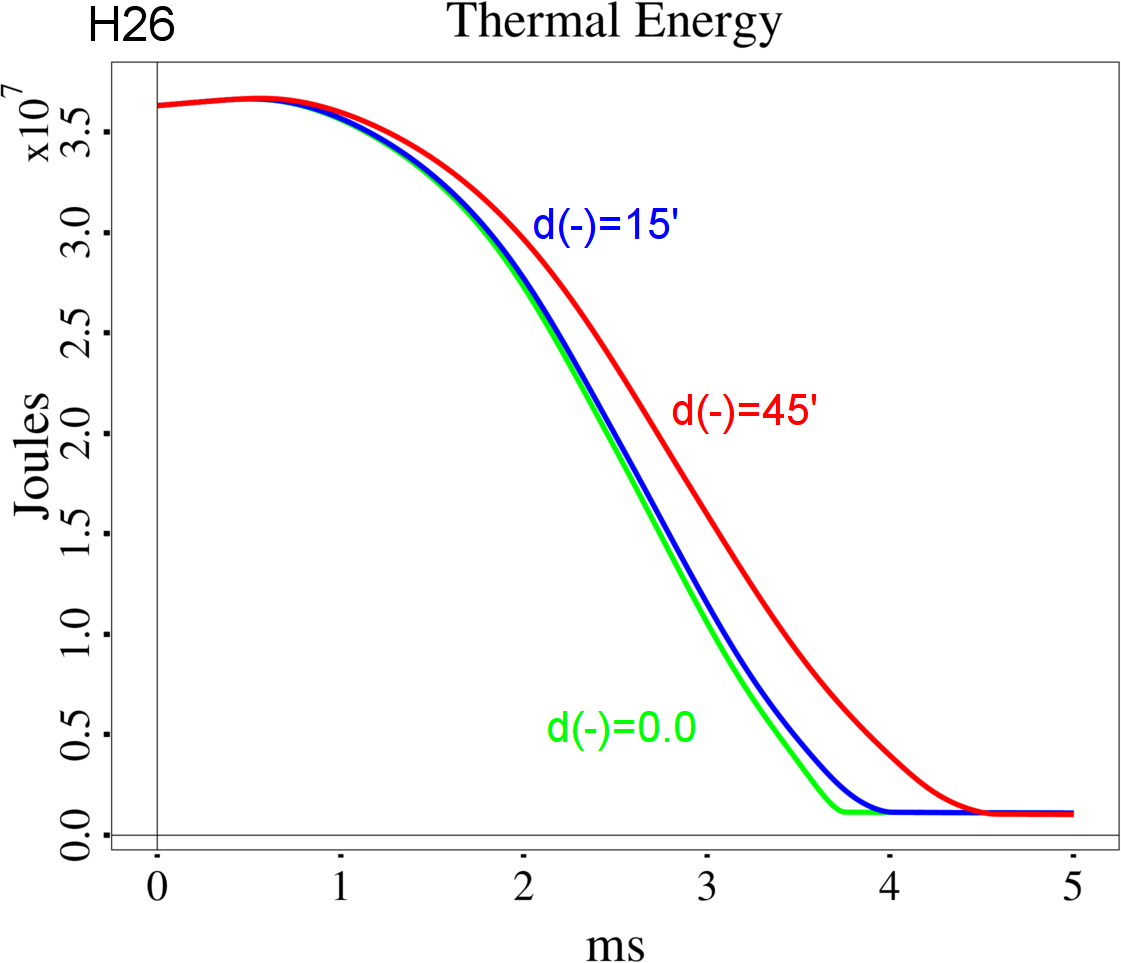}
		\caption{H26 L-mode}
		\label{fig:H26_TES4}
	\end{subfigure}
	\hspace{0.00cm}
	\begin{subfigure}[b]{0.375\textwidth}
		\includegraphics[width=1.0\textwidth]{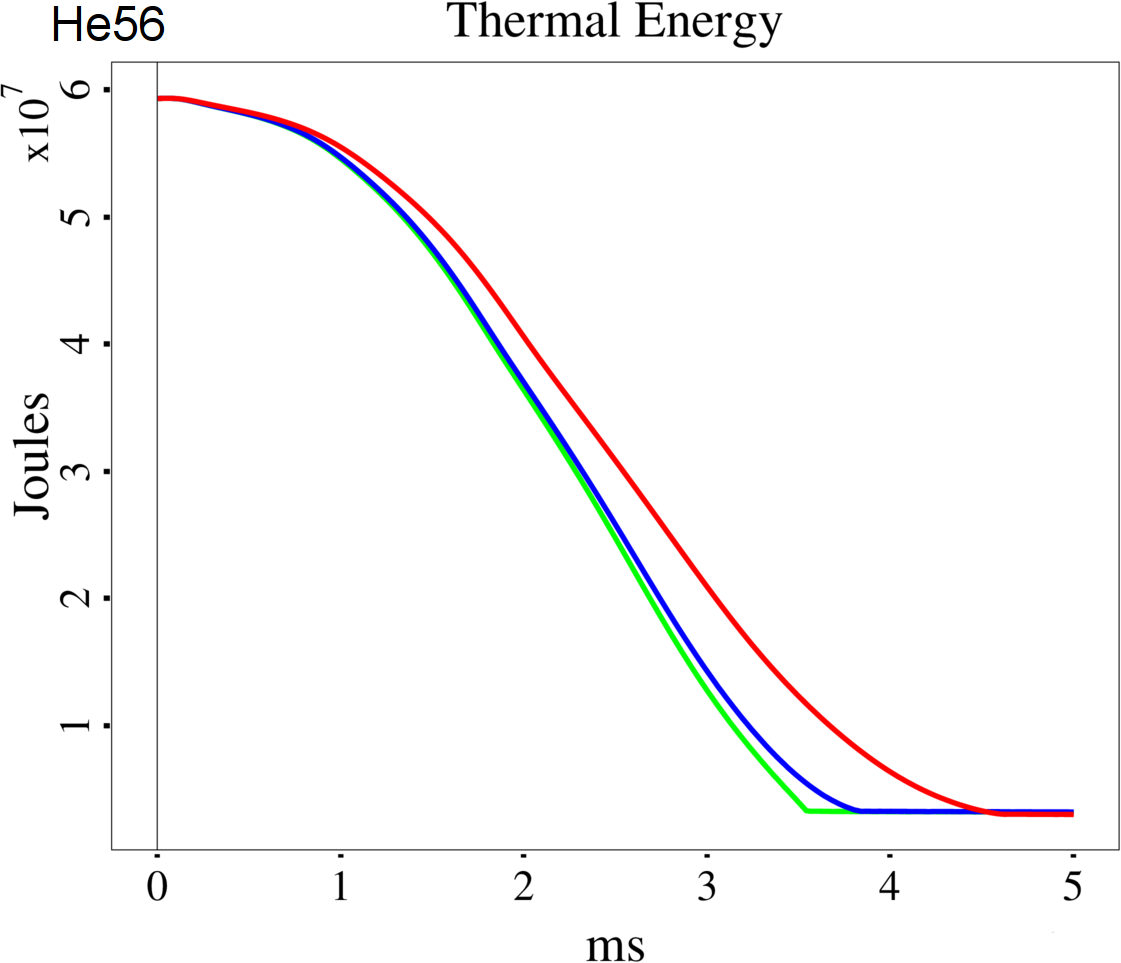}
		\caption{He56}
		\label{fig:He56_TES4}
	\end{subfigure}
	\hspace{0.00cm}
	\begin{subfigure}[b]{0.375\textwidth}
		\includegraphics[width=1.0\textwidth]{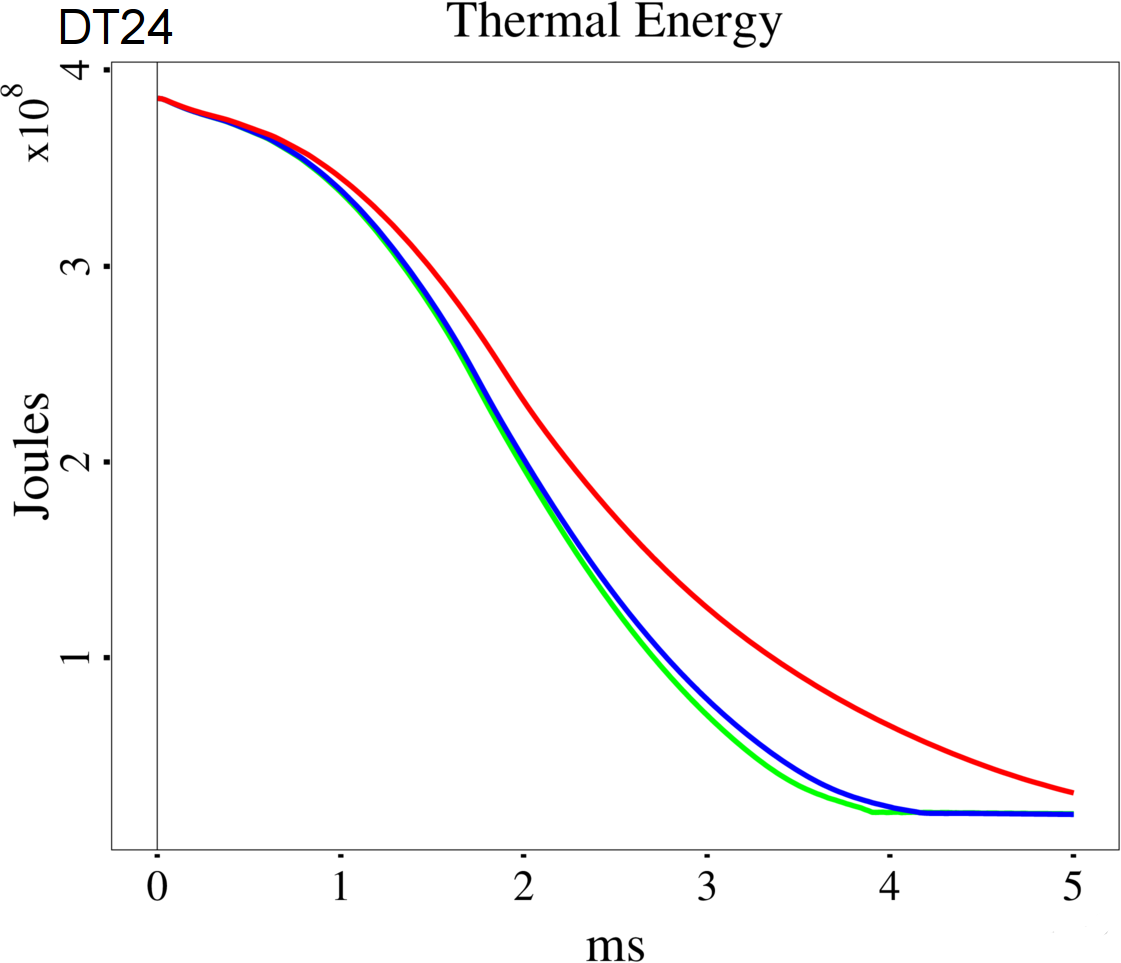}
		\caption{DT24}
		\label{fig:DT24_TES4}
	\end{subfigure}
}
	\caption{Comparison of thermal energy for {\bf S4} Poloidal Extent shows slower thermal quench for larger angular
	dispersion.  DT24 (\ref{fig:DT24_TES4}) d$\theta$=45$^{\circ}$ shows incomplete quench.}
	\label{fig:TES4}
\end{figure}

For {\bf S4} Poloidal Extent, fragment injection angles are randomly assigned from a uniform distribution between
$\theta$=[-d$\theta_{hw}$,+d$\theta_{hw}$].  Fragment velocities  are randomly assigned from a uniform distribution between
v=[v$_0$-dv,v$_0$+dv] as in Scenario 3.  As with Scenario 3, the uniform distribution is chosen over the more
physical Gaussian distribution to better assess the impact of the poloidal spread.

The scan in poloidal extent shows that a wider poloidal extent slows the thermal quench by reducing the effective
velocity of the fragments by $\sim\cos(\Delta\theta)$.  Some fragments with the largest poloidal injection angles
may even miss the plasma as the cross section decreases due to the progress of the thermal quench.

Figures~\ref{fig:TES4} show a comparison of the thermal energy for the two poloidal dispersions
d$\theta_{hw}$=[\clrb{15$^{\circ}$},\clrr{45$^{\circ}$}].  The dv/v=0.2 case with \clrlg{d$\theta_{hw}$=0$^{\circ}$}
from Scenario 3 is included for reference.  The plots show that \clrb{d$\theta_{hw}$=15$^{\circ}$} closely tracks the
\clrlg{d$\theta_{hw}$=0$^{\circ}$}.  A Gaussian spread would likely overlay the \clrlg{d$\theta_{hw}$=0$^{\circ}$} plot.  

Tables~\ref{tab:s4tqm} list the thermal quench times and assimilation and radiation fractions.  Again, we see how
closely the \clrb{d$\theta_{hw}$=15$^{\circ}$} tracks the \clrlg{d$\theta_{hw}$=0$^{\circ}$}.  Not surprisingly, the wider
poloidal extend decreases assimilation.

The DT24 (\ref{fig:DT24_TES4}) \clrr{d$\theta_{hw}$=45$^{\circ}$} case shows an incomplete quench (recall this is the
larger r=25.0mm pellet S=10 plume).  Table~\ref{tab:s4DT24tqm} indicates that the incomplete quench is due to $\sim1/3$
of the fragments missing the plasma.  This is inferred from the $\sim2/3$ assimilated fragments.

\section{S5 - Poloidal Injection Angle : $\theta$ = $\pm$[\clrc{2}\clrb{0}$^{\circ}$,\clrm{4}\clrr{5}$^{\circ}$] (dv/v=0.2)}

\begin{table}
	\begin{subtable}{1.0\textwidth}
\centerline{
\begin{tabular}{|r||c|c|c|}\hline
	H26 scenario 5		&\clrlg{$\theta$=0$^{\circ}$}&\clrc{+20$^{\circ}$}/\clrb{-20$^{\circ}$}&\clrm{+45$^{\circ}$}/\clrr{-45$^{\circ}$}\\ \hline \hline
	($\tau_{TQ}$),T.R.	&       (3.76ms),0.03        &           0.29/0.18                     &           0.86/0.77                     \\ \hline
	assim.			&        0.22                &           0.20/0.21                     &           0.10/0.12                     \\ \hline
	rad. frac.		&        0.97                &           1.00/0.99                     &           1.41/1.27                     \\ \hline
	q=2 $\tau_{TQ}$(ms)	&        1.18                &           1.26/1.24                     &           1.71/1.59                     \\ \hline
	q=2 assim.              &        0.023               &           0.034/0.034                   &           0.032/0.032                   \\ \hline
\end{tabular}
	}
	\caption{H26 L-mode}
	\label{tab:s5h26ltqm}
\end{subtable}

	\vspace{0.25cm}
	\begin{subtable}{1.0\textwidth}
\centerline{
\begin{tabular}{|r||c|c|c|}\hline
	He56 scenario 5		&\clrlg{$\theta$=0$^{\circ}$}&\clrc{+20$^{\circ}$}/\clrb{-20$^{\circ}$}&\clrm{+45$^{\circ}$}/\clrr{-45$^{\circ}$}\\ \hline \hline
	($\tau_{TQ}$),T.R.	&       (3.56ms),0.06        &           0.26/0.20                     &           0.74/0.68                     \\ \hline
	assim.			&       0.34                 &           0.31/0.32                     &           0.19/0.22                     \\ \hline
	rad. frac.		&       0.86                 &           0.99/0.98                     &           1.37/1.29                     \\ \hline
	q=2 $\tau_{TQ}$(ms)	&       1.25                 &           1.33/1.30                     &           1.77/1.66                     \\ \hline
	q=2 assim.         	&       0.109                &           0.108/0.106                   &           0.105/0.106                    \\ \hline
\end{tabular}
	}
	\caption{He56}
	\label{tab:s5he56tqm}
\end{subtable}

	\vspace{0.25cm}
	\begin{subtable}{1.0\textwidth}
\centerline{
\begin{tabular}{|r||c|c|c|}\hline
	DT24 scenario 5		&\clrlg{$\theta$=0$^{\circ}$}&\clrc{+20$^{\circ}$}/\clrb{-20$^{\circ}$}&\clrm{+45$^{\circ}$}/\clrr{-45$^{\circ}$}\\ \hline \hline
	($\tau_{TQ}$),T.R.	&       (3.90ms),0.05        &           0.18/0.15                     &           0.64/0.58                     \\ \hline
	assim.			&       0.73                 &           0.70/0.70                     &           0.49/0.52                     \\ \hline
	rad. frac.		&       0.85                 &           0.92/0.90                     &           1.11/1.09                     \\ \hline
	q=2 $\tau_{TQ}$(ms) 	&       1.44                 &           1.51/1.49                     &           1.91/1.83                     \\ \hline
	q=2 assim.          	&       0.375                &           0.367/0.366                   &           0.330/0.338                   \\ \hline
\end{tabular}
	}
	\caption{DT24}
	\label{tab:s5dt24tqm}
\end{subtable}
	\caption{Thermal Quench Metrics for {\bf S5} Poloidal Injection Angle.   The thermal quench time for the pencil
	beam case is used to measure the metrics since none achieve complete thermal quench.  Listed are the thermal
	remnant (T.R.), assimilation and radiation fractions at t=$\tau_{TQ}$ and q=2 quench time and assimilation.  Large injection angles reduce the
	thermal quench efficiency.  \clrr{$\theta$=$\pm$45$^{\circ}$} radiation fractions >1 are a result of incomplete
	accounting of viscous heating.}
	\label{tab:s5tqm}
\end{table}

\begin{figure}
\centerline{
	\hspace{0.00cm}
	\begin{subfigure}[b]{0.375\textwidth}
		\includegraphics[width=1.0\textwidth]{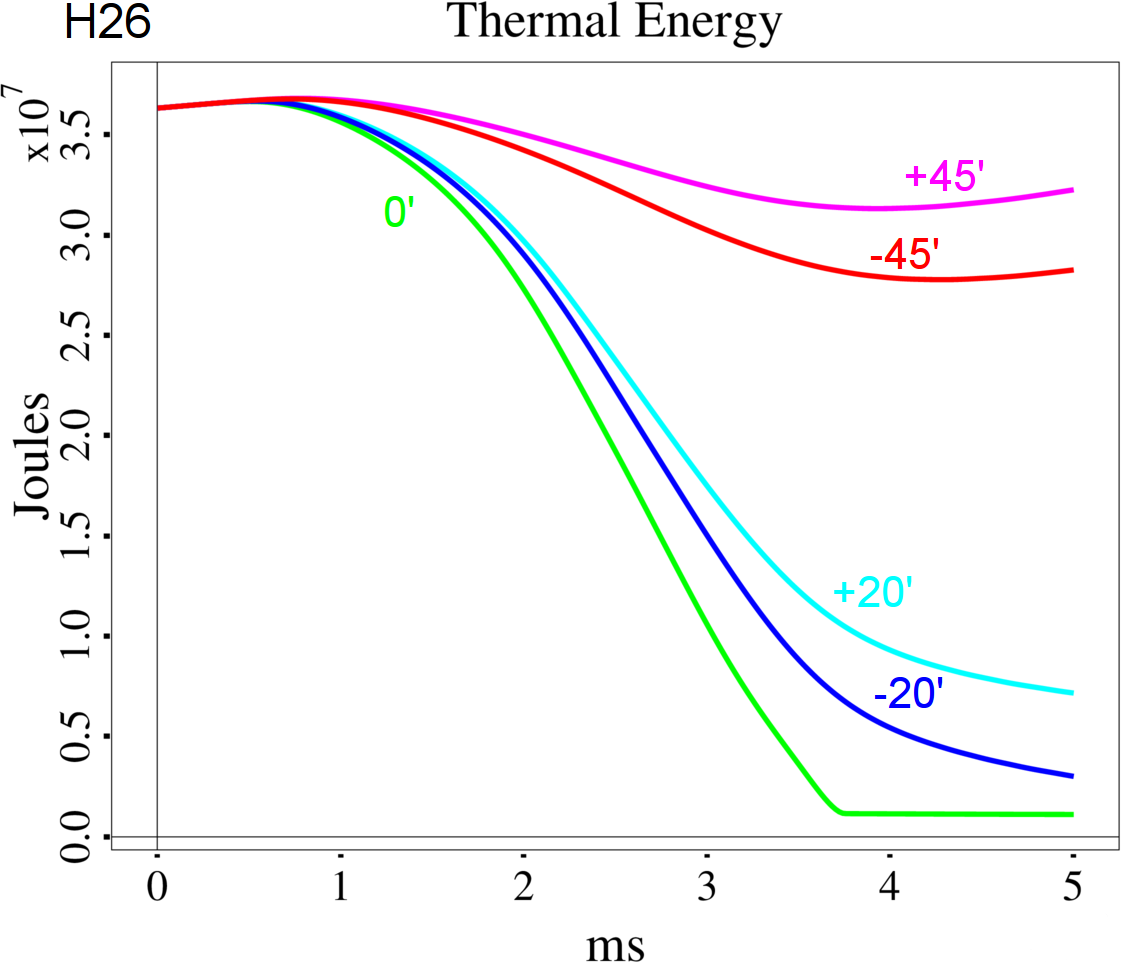}
		\caption{H26 L-mode}
		\label{fig:H26_TES5}
	\end{subfigure}
	\hspace{0.00cm}
	\begin{subfigure}[b]{0.375\textwidth}
		\includegraphics[width=1.0\textwidth]{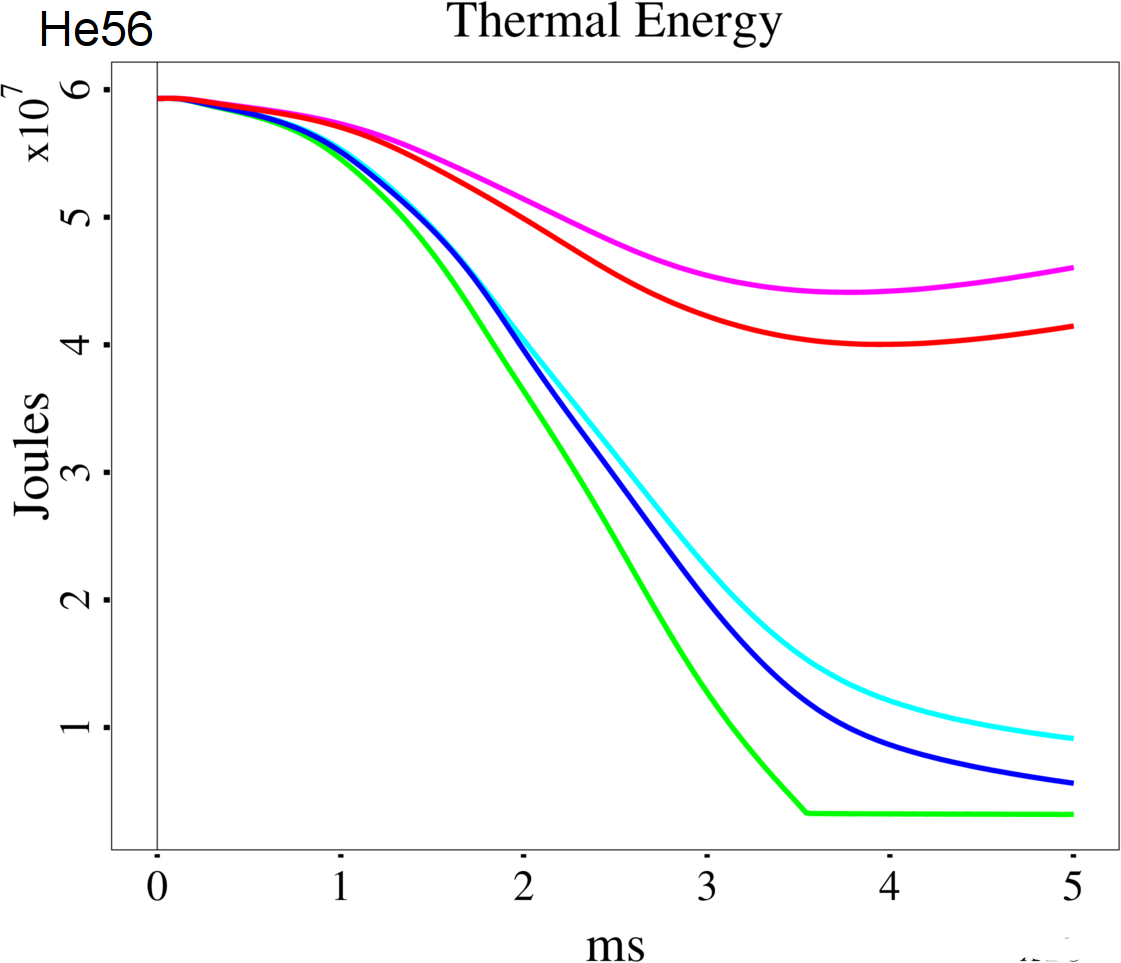}
		\caption{He56}
		\label{fig:He56_TES5}
	\end{subfigure}
	\hspace{0.00cm}
	\begin{subfigure}[b]{0.375\textwidth}
		\includegraphics[width=1.0\textwidth]{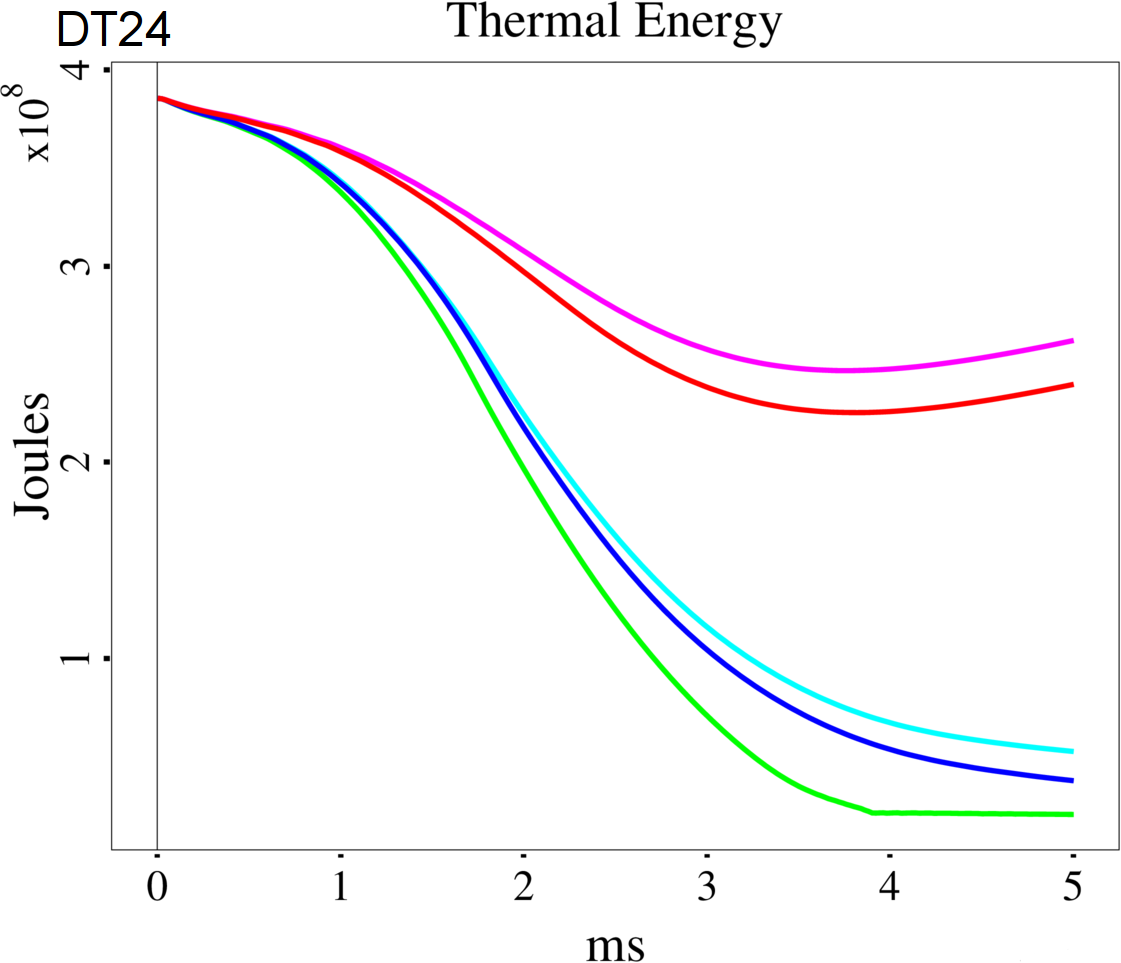}
		\caption{DT24}
		\label{fig:DT24_TES5}
	\end{subfigure}
}
	\caption{Comparison of thermal energy for {\bf S5} Poloidal Injection Angle shows less complete quench for
	larger poloidal injection angles.  Asymmetry in $\pm$ injection angles due to offset of injector above magnetic
	axis resulting in negative angles intersecting more of the plasma.}
	\label{fig:TES5}
\end{figure}

The {\bf S5} Poloidal Injection Angle scan uses fragments of radius r=2.5mm, with velocity v=500m/s and a velocity
dispersion of dv/v=0.2.  All cases result in an incomplete thermal quench due to the tangential trajectory with respect
to the plasma core.

Figures~\ref{fig:TES5} show the thermal energies for the four injection angles 
$\theta$=[\clrb{-20$^{\circ}$},\clrc{+20$^{\circ}$},\clrr{-45$^{\circ}$},\clrm{+45$^{\circ}$}].  The negative injection
angles result in more thermal quenching due to the slight offset above the magnetic axis of equatorial SPI injector
EQ\_08\_4\_1 (fig.~\ref{fig:spiinjector}).  The offset causes negative angle trajectories to intersect a little more
of the plasma core and more quenching.  An injector aligned with the magnetic axis is expected to produce more symmetric
results.

Tables~\ref{tab:s5tqm} list the thermal quench metrics for {\bf S5}.  Since all cases result in an incomplete thermal
quench, we include the thermal remnant (residual thermal energy/initial thermal energy) and assimilation and radiation
fractions as measured at the thermal quench time of the reference \clrlg{$\theta$=0$^{\circ}$} case.  This thermal
quench time is listed in tables~\ref{tab:s5tqm}.

Despite the tangential trajectory and incomplete quench, the \clrb{$\theta$=$\pm$20$^{\circ}$} cases measure
surprisingly close to the \clrlg{$\theta$=0$^{\circ}$} case.  Not surprisingly, the larger poloidal injection angles
result in lower thermal quench and lower assimilation.  

%\subsection{Flow Generation From Injected Fragments (H26, $\theta_{inj}$=0$^{\circ}$,$\pm$20$^{\circ}$, t=5.0ms)}
\subsection{Flow Generation From Injected Fragments}

\begin{figure}
\centerline{
	\hspace{0.00cm}
	\begin{subfigure}[b]{0.375\textwidth}
		\includegraphics[width=1.00\textwidth]{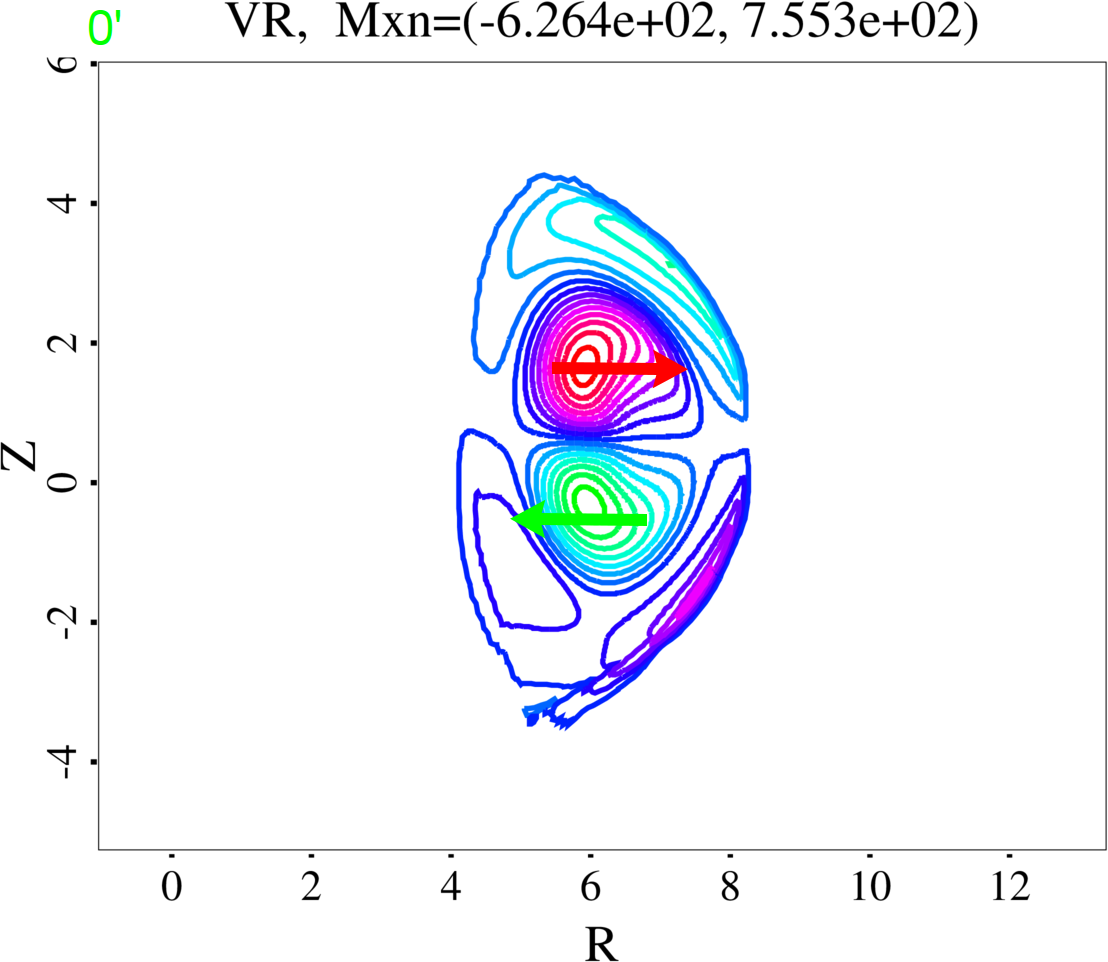}
		\caption{$\theta$=0$^{\circ}$}
		\label{fig:vpol_0}
	\end{subfigure}
	\hspace{0.00cm}
	\begin{subfigure}[b]{0.375\textwidth}
		\includegraphics[width=1.00\textwidth]{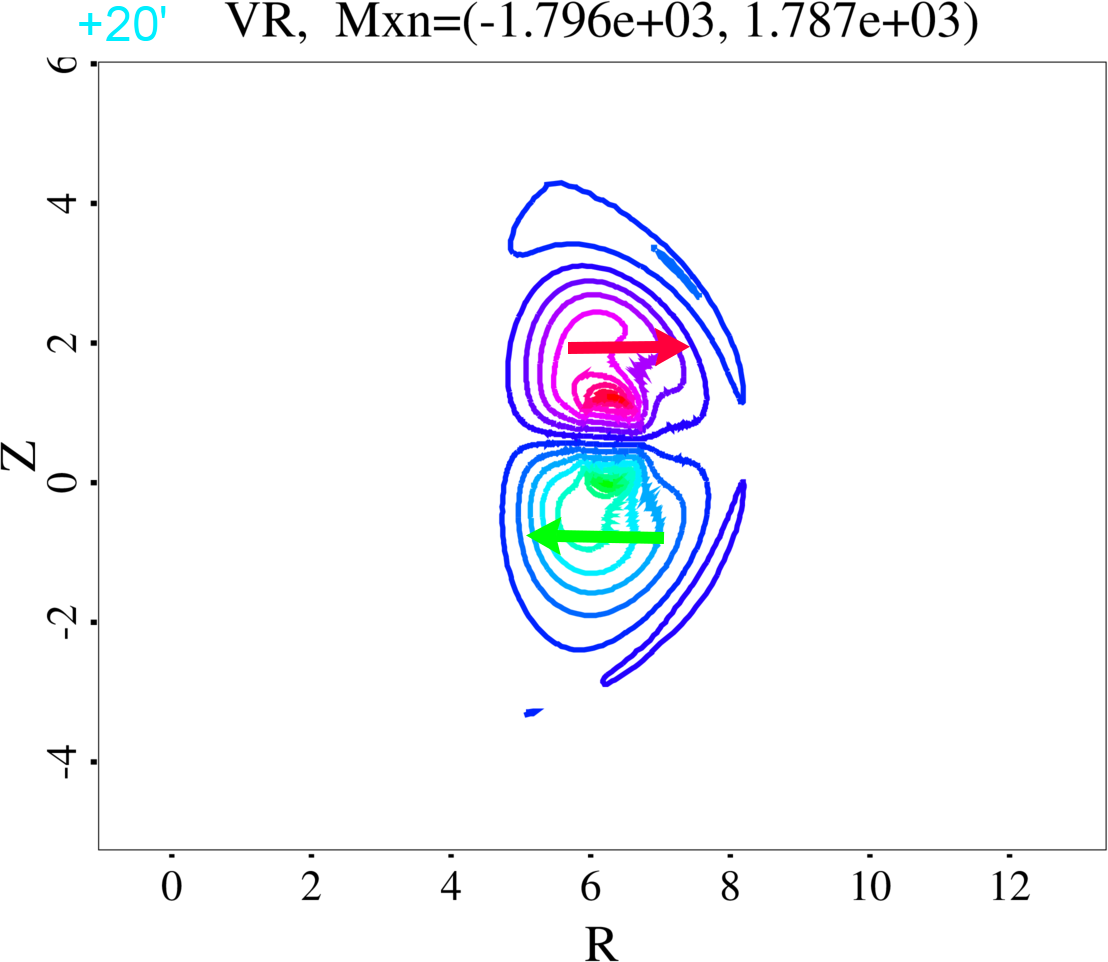}
		\caption{$\theta$=+20$^{\circ}$}
		\label{fig:vpol_p20}
	\end{subfigure}
	\hspace{0.00cm}
	\begin{subfigure}[b]{0.375\textwidth}
		\includegraphics[width=1.00\textwidth]{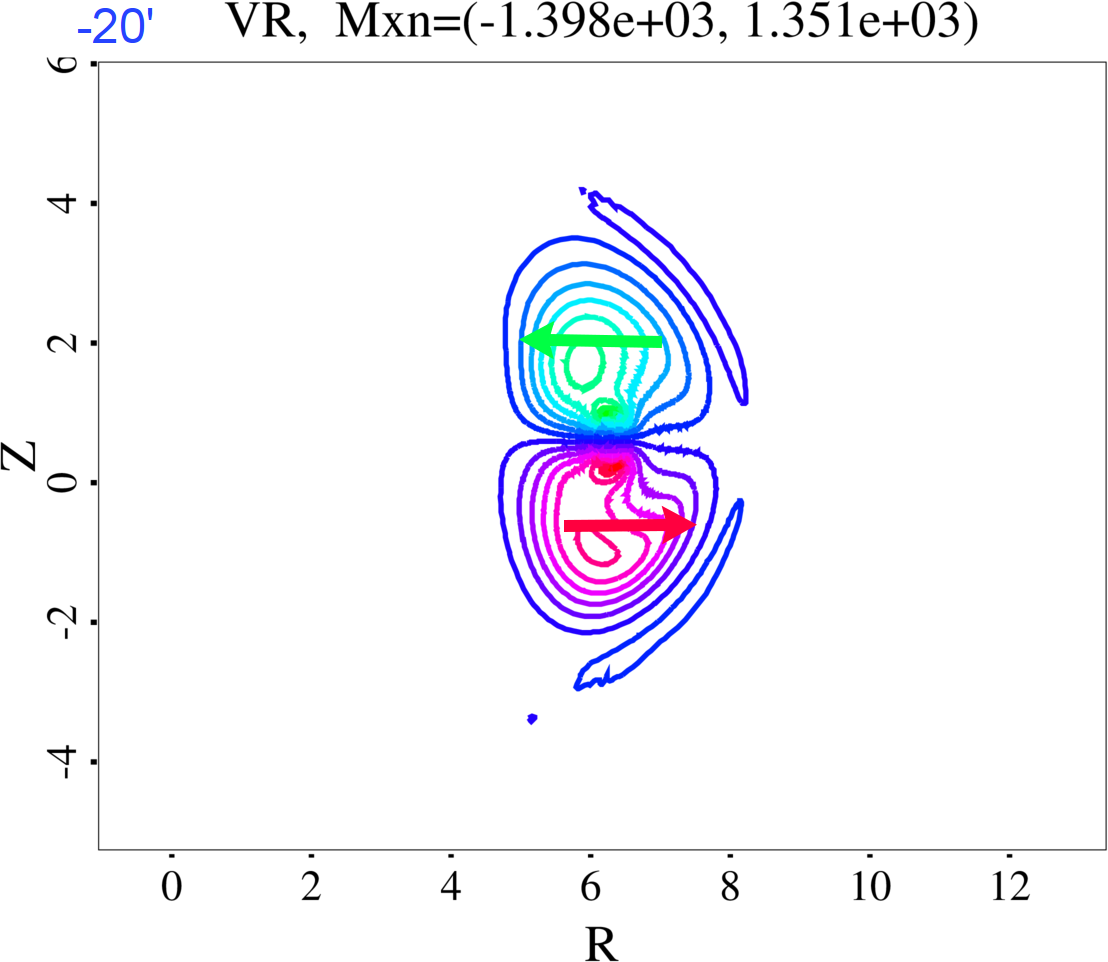}
		\caption{$\theta$=-20$^{\circ}$}
		\label{fig:vpol_n20}
	\end{subfigure}
}
	\vspace{0.25cm}
\centerline{
	\hspace{0.00cm}
	\begin{subfigure}[b]{0.375\textwidth}
		\includegraphics[width=1.00\textwidth]{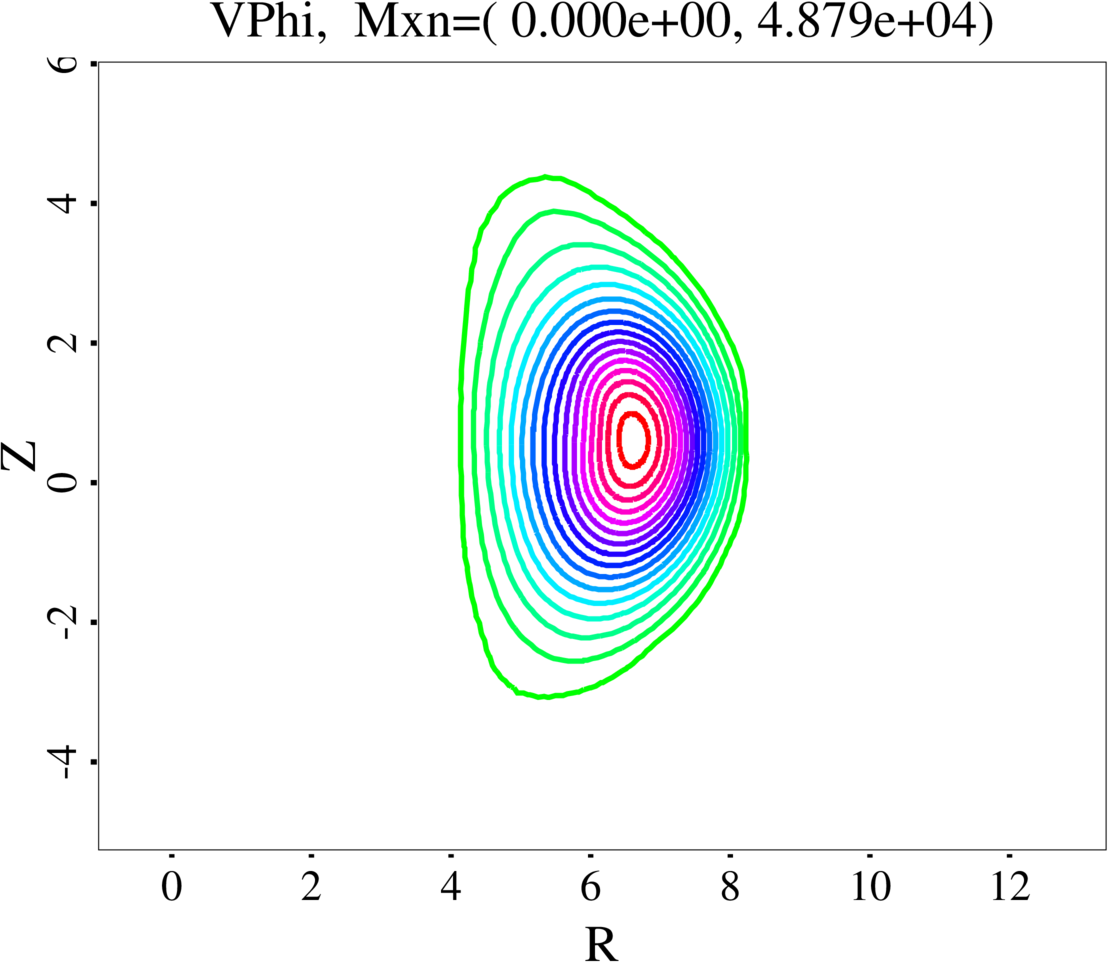}
		\caption{$\theta$=0$^{\circ}$}
		\label{fig:vtor_0}
	\end{subfigure}
	\hspace{0.00cm}
	\begin{subfigure}[b]{0.375\textwidth}
		\includegraphics[width=1.00\textwidth]{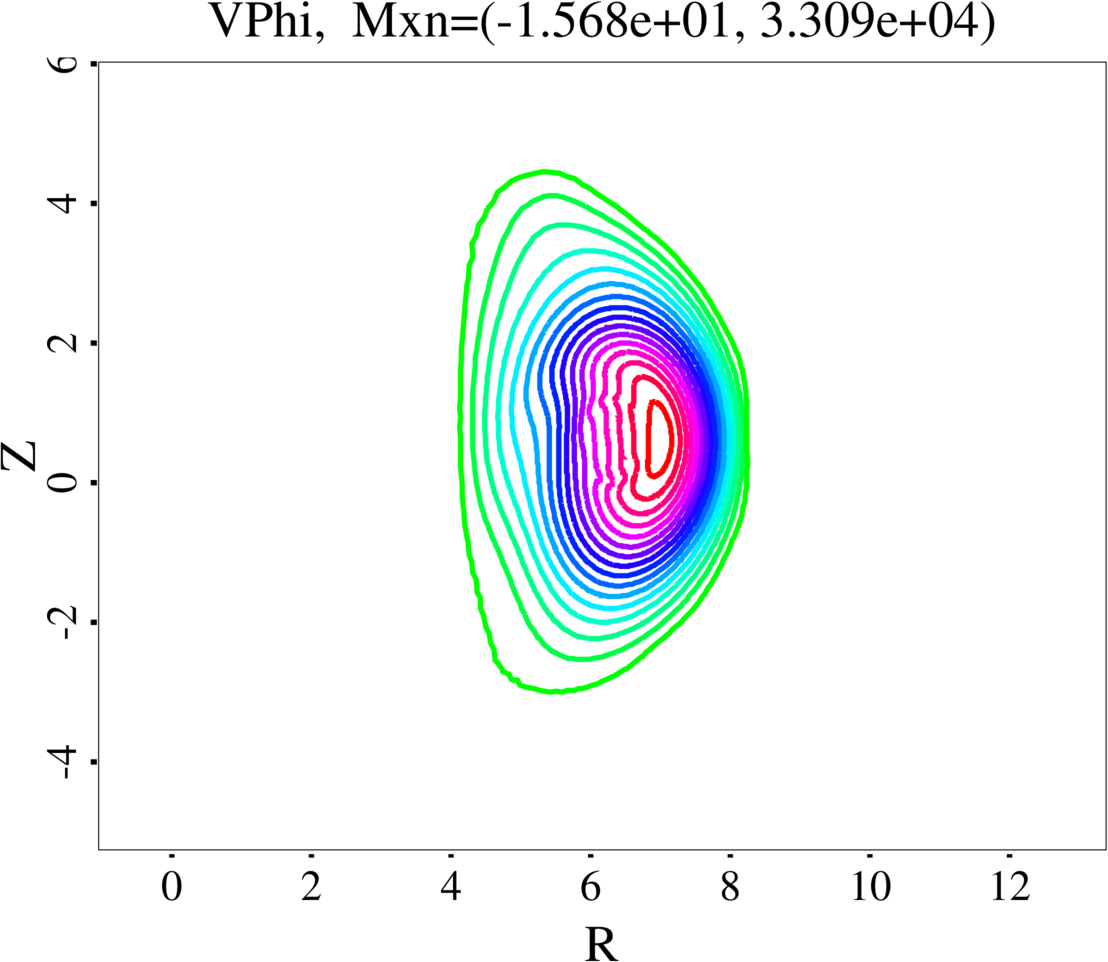}
		\caption{$\theta$=+20$^{\circ}$}
		\label{fig:vtor_p20}
	\end{subfigure}
	\hspace{0.00cm}
	\begin{subfigure}[b]{0.375\textwidth}
		\includegraphics[width=1.00\textwidth]{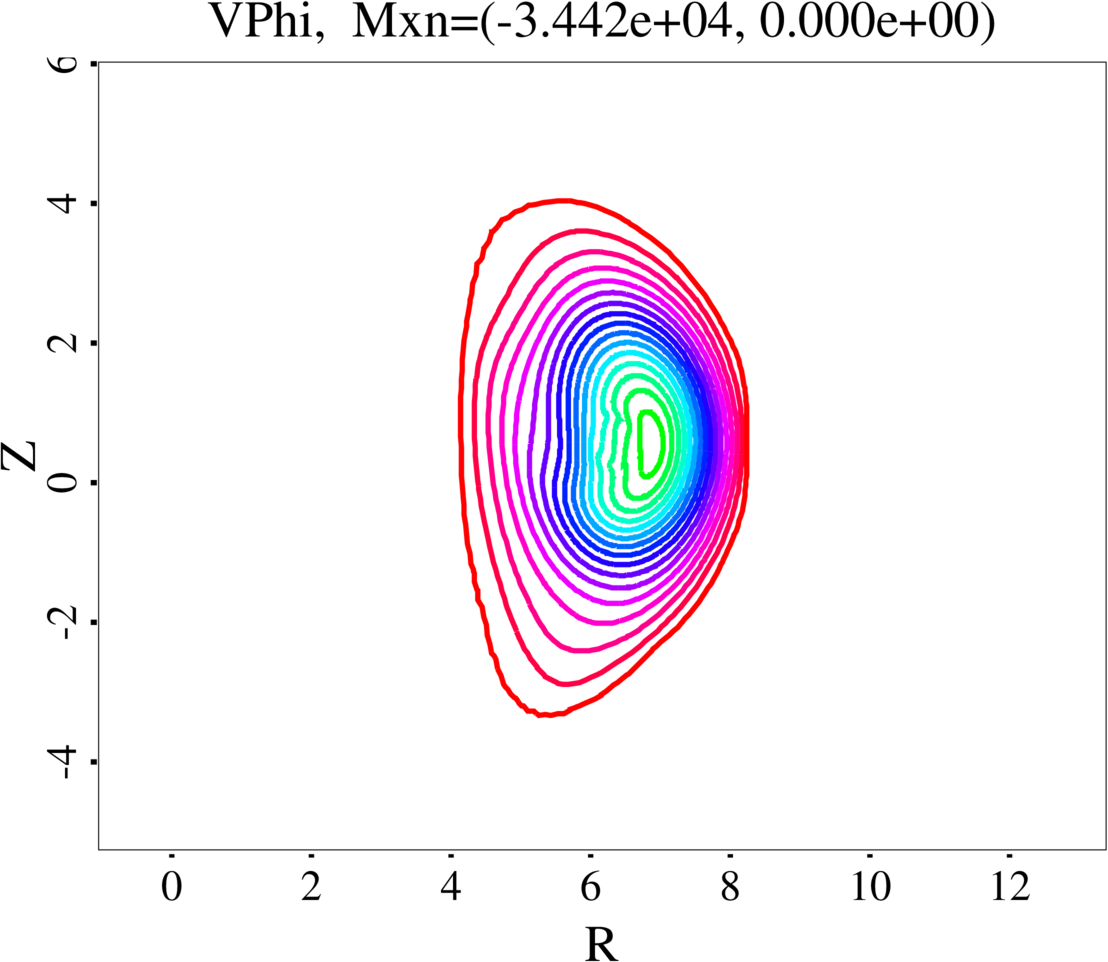}
		\caption{$\theta$=-20$^{\circ}$}
		\label{fig:vtor_n20}
	\end{subfigure}
}
	\caption{Poloidal (V$_R$) flow contours (\ref{fig:vpol_0},\ref{fig:vpol_p20},\ref{fig:vpol_n20}) for H26 L-mode
	at t=5.0ms shows flows generated by injection angles = [$\theta_{inj}$=0$^{\circ}$,$\pm$20$^{\circ}$] are counter to
	the direction of fragment injection.  The toroidal flow (\ref{fig:vtor_0},\ref{fig:vtor_p20},\ref{fig:vtor_n20})
	also changes direction.} 
	\label{fig:VS5}
\end{figure}

It was observed in the {\bf S5} scans that the flow changes direction with the sign of the injection angle.
Figures~\ref{fig:VS5} show the radial and toroidal velocity components at t=5.0ms for equilibrium H26,
$\theta_{inj}$=[\clrlg{0$^{\circ}$},\clrb{-20$^{\circ}$},\clrc{+20$^{\circ}$}].  These contours show that both the
poloidal and toroidal flow directions change with opposite signs of injection angle.  The contours also show that the flow is
opposite to the direction of the fragment velocities.  The plasma flow is driven by the asymmetric flow generated by the ablation
of the fragments.  The ablation driven flow flows along the magnetic field.  The outboard directed momentum dominates and
dictates the direction of the global flow.  Comparison with the \clrlg{$\theta_{inj}$=0$^{\circ}$} shows that finite
injection angles increase the poloidal flow but decrease the toroidal flow.  

The larger than typical poloidal flows driven in Scenario 5 are the source of the >1 radiation fractions listed in 
tables~\ref{tab:s5tqm}.  As mentioned in the introduction, the viscous heating is not kept track of and kinetic energy is
used in its place in computing the radiation fraction which leads to inaccurate radiation fractions, particularly in cases where 
flows are strong and viscous heating is a larger component.

These global flows may only be an axisymmetric phenomenon and not survive into 3D.  However, if they do so, two
questions arise: are there benefits to driving flow with SPI? do the flows exert a force on the fragments and alter
their trajectories?

\section{Deposition Radius and Offset}

To complement the fragment plume parameter scans of Scenarios 1-5, we present a scan of deposition parameters: the
deposition radius and deposition offset.  Recall that the SPI algorithm\cite{kimc:2019} deposits the ablated neutrals as
a Gaussian circle of radius r$_d$ in the poloidal plane and as a vonMises distribution (periodic Gaussian) in the
toroidal plane.  For all scenarios presented above, the deposition radius is 30.0cm.  The center of deposition can
optionally be offset from the center of field (n,T) evaluation, depositing the neutrals either behind(r$_d$>0) or in
front of(r$_d$<0) the center of evaluation.  The default is to co-locate the center of deposition and evaluation.  

\subsection{Deposition Radius}

\begin{table}
	\begin{subtable}{1.0\textwidth}
\centerline{
\begin{tabular}{|r||c|c|c|}\hline
	H26 deposition radius   &\clrlg{15.0cm}&\clrb{20.0cm}&\clrr{30.0cm}\\ \hline \hline
	$\tau_{TQ}$(ms)		&     3.78     &     3.80    &     3.76    \\ \hline
	assim.			&     0.206    &     0.210   &     0.218   \\ \hline
	rad. frac.		&     1.02     &     1.00    &     0.97    \\ \hline
	q=2 $\tau_{TQ}$(ms)	&     1.09     &     1.12    &     1.18    \\ \hline
	q=2 assim.         	&     0.019    &     0.023   &     0.023   \\ \hline
\end{tabular}
	}
	\caption{H26 L-mode}
	\label{tab:H26l_dr}
\end{subtable}

	\vspace{0.25cm}
	\begin{subtable}{1.0\textwidth}
\centerline{
\begin{tabular}{|r||c|c|c|}\hline
	DT24 deposition radius  &\clrlg{20.0cm}&\clrb{30.0cm}&\clrr{45.0cm}\\ \hline \hline
	$\tau_{TQ}$(ms)		&     3.80     &     3.90    &     3.80    \\ \hline
	assim.			&     0.717    &     0.726   &     0.734   \\ \hline
	rad. frac.		&     0.89     &     0.85    &     0.82    \\ \hline
	q=2 $\tau_{TQ}$(ms)	&     1.32     &     1.44    &     1.50    \\ \hline
	q=2 assim.         	&     0.293    &     0.375   &     0.450   \\ \hline
\end{tabular}
	}
	\caption{DT24}
	\label{tab:DT24_dr}
\end{subtable}
	\caption{Thermal Quench Metrics for deposition radius listing thermal quench time($\tau_{TQ}$),
	assimilation and radiation fractions at t=$\tau_{TQ}$ and q=2 quench time and assimilation shows similar quench times and modest
	increase in assimilation but decrease in radiation.}
	\label{tab:depr}
\end{table}

\begin{figure}
\centerline{
	\hspace{0.00cm}
	\begin{subfigure}[b]{0.50\textwidth}
		\includegraphics[width=1.00\textwidth]{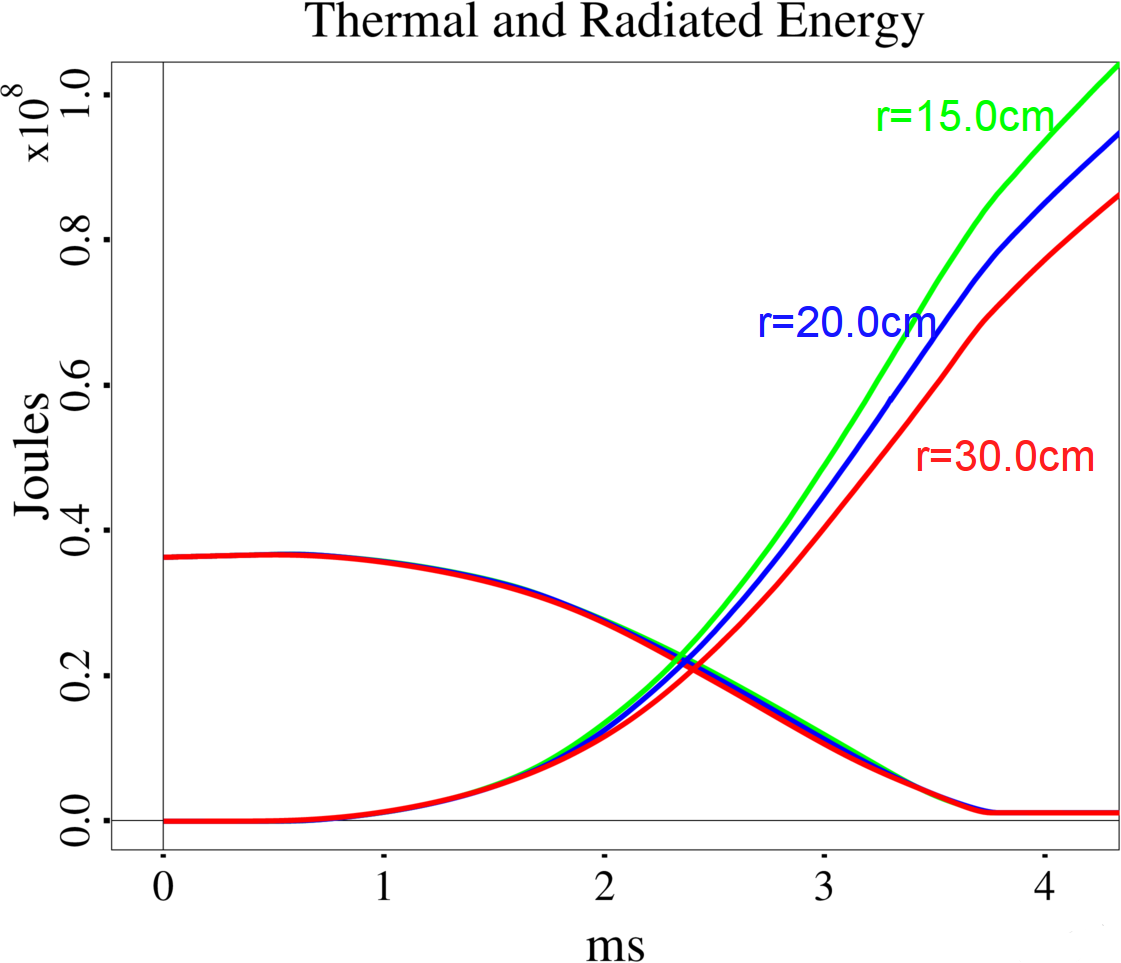}
		\caption{H26 L-mode}
		\label{fig:H26l_depsize}
	\end{subfigure}
	\hspace{0.00cm}
	\begin{subfigure}[b]{0.50\textwidth}
		\includegraphics[width=1.00\textwidth]{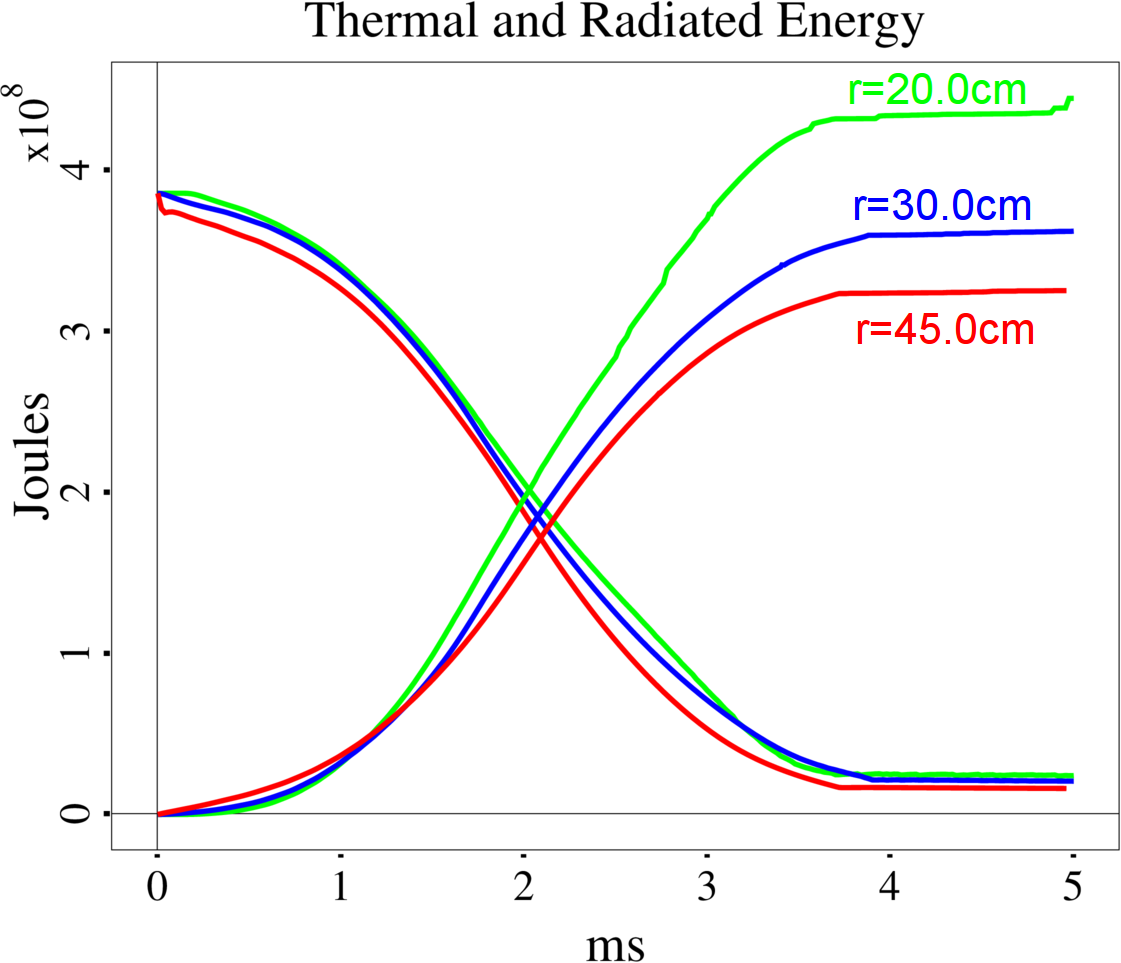}
		\caption{DT24}
		\label{fig:DT24_depsize}
	\end{subfigure}
}
	\caption{Comparison of thermal and radiated energies for various deposition radii for
	H26(\ref{fig:H26l_depsize}) and DT24(\ref{fig:DT24_depsize}) show modest impact of changing deposition footprint
	for these axisymmetric cases.}
	\label{fig:depsize}
\end{figure}

Figures~\ref{fig:depsize} show a comparison of the thermal and radiated energies for various deposition radii for 
H26(\ref{fig:H26l_depsize}) and DT24(\ref{fig:DT24_depsize}).  Deposition radius does not have a significant impact on
the thermal energy evolution.  The modest difference seen in the high thermal energy content equilibrium DT24 can be
attributed more to burn through of the fragment plume delaying the thermal quench time.  

Tables~\ref{tab:depr} of the usual metrics show a modest increase in assimilation and decrease in the radiation
fraction as the deposition radius is increased.  The earlier q=2 quench times can be attributed directly to the geometry
of the larger deposition radius.

\subsection{Deposition Offset}

\begin{table}
	\begin{subtable}{1.0\textwidth}
\centerline{
\begin{tabular}{|r||c|c|c|}\hline
	H26 deposition offset   &\clrlg{0.0cm}&\clrb{15.0cm}&\clrr{30.0cm}\\ \hline \hline
	$\tau_{TQ}$(ms)		&      4.24    &     4.24    &     4.24   \\ \hline
	assim.			&      0.223   &     0.326   &     0.517  \\ \hline
	rad. frac.		&      0.97    &     0.97    &     0.90   \\ \hline
	q=2 $\tau_{TQ}$(ms)	&      1.49    &     1.43    &     1.42   \\ \hline
	q=2 assim.         	&      0.043   &     0.038   &     0.051  \\ \hline
\end{tabular}
	}
	\caption{H26 L-mode}
	\label{tab:H26_do}
\end{subtable}

	\vspace{0.25cm}
	\begin{subtable}{1.0\textwidth}
\centerline{
\begin{tabular}{|r||c|c|c|}\hline
	DT24 deposition offset  &\clrlg{0.0cm}&\clrb{10.0cm}&\clrr{20.0cm}\\ \hline \hline
	$\tau_{TQ}$(ms)		&      4.30    &     4.30    &     4.30    \\ \hline
	assim.			&      0.718   &     0.912   &     1.000   \\ \hline
	rad. frac.		&      0.87    &     0.90    &     0.75    \\ \hline
	q=2 $\tau_{TQ}$(ms)	&      1.87    &     1.79    &     1.76    \\ \hline
	q=2 assim.         	&      0.367   &     0.408   &     0.483   \\ \hline
\end{tabular}
	}
	\caption{DT24}
	\label{tab:DT24_do}
\end{subtable}
	\caption{Thermal Quench Metrics for deposition offset listing thermal quench time($\tau_{TQ}$),
	assimilation and radiation fractions at t=$\tau_{TQ}$ and q=2 quench time and assimilation shows similar quench times and
	increasing assimilation but modest change in radiation.}
	\label{tab:depo}
\end{table}

\begin{figure}
\centerline{
	\hspace{0.00cm}
	\begin{subfigure}[b]{0.50\textwidth}
		\includegraphics[width=1.00\textwidth]{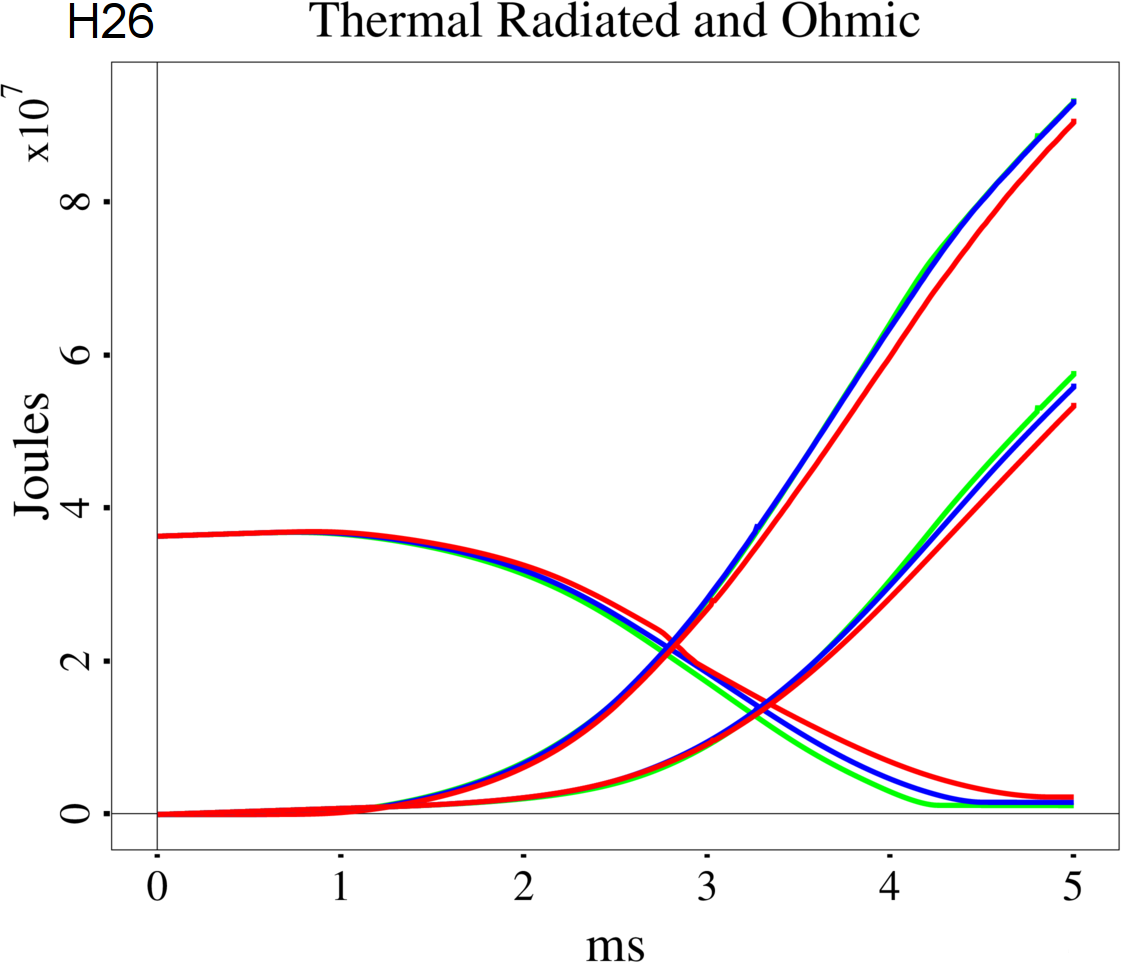}
		\caption{H26 L-mode}
		\label{fig:H26l_Toff}
	\end{subfigure}
	\hspace{0.00cm}
	\begin{subfigure}[b]{0.50\textwidth}
		\includegraphics[width=1.00\textwidth]{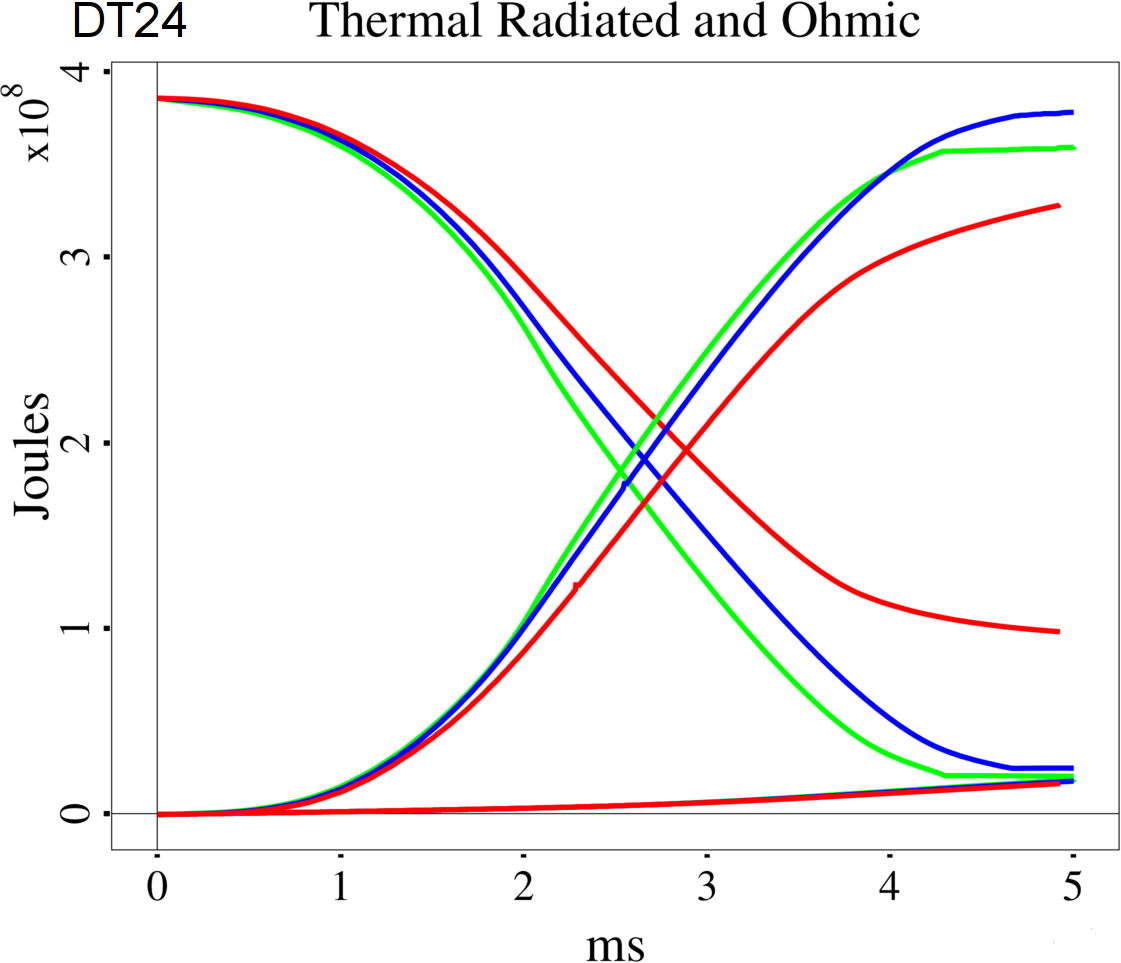}
		\caption{DT24}
		\label{fig:DT24_Toff}
	\end{subfigure}
}
	\vspace{0.25cm}
\centerline{
	\hspace{0.00cm}
	\begin{subfigure}[b]{0.50\textwidth}
		\includegraphics[width=1.00\textwidth]{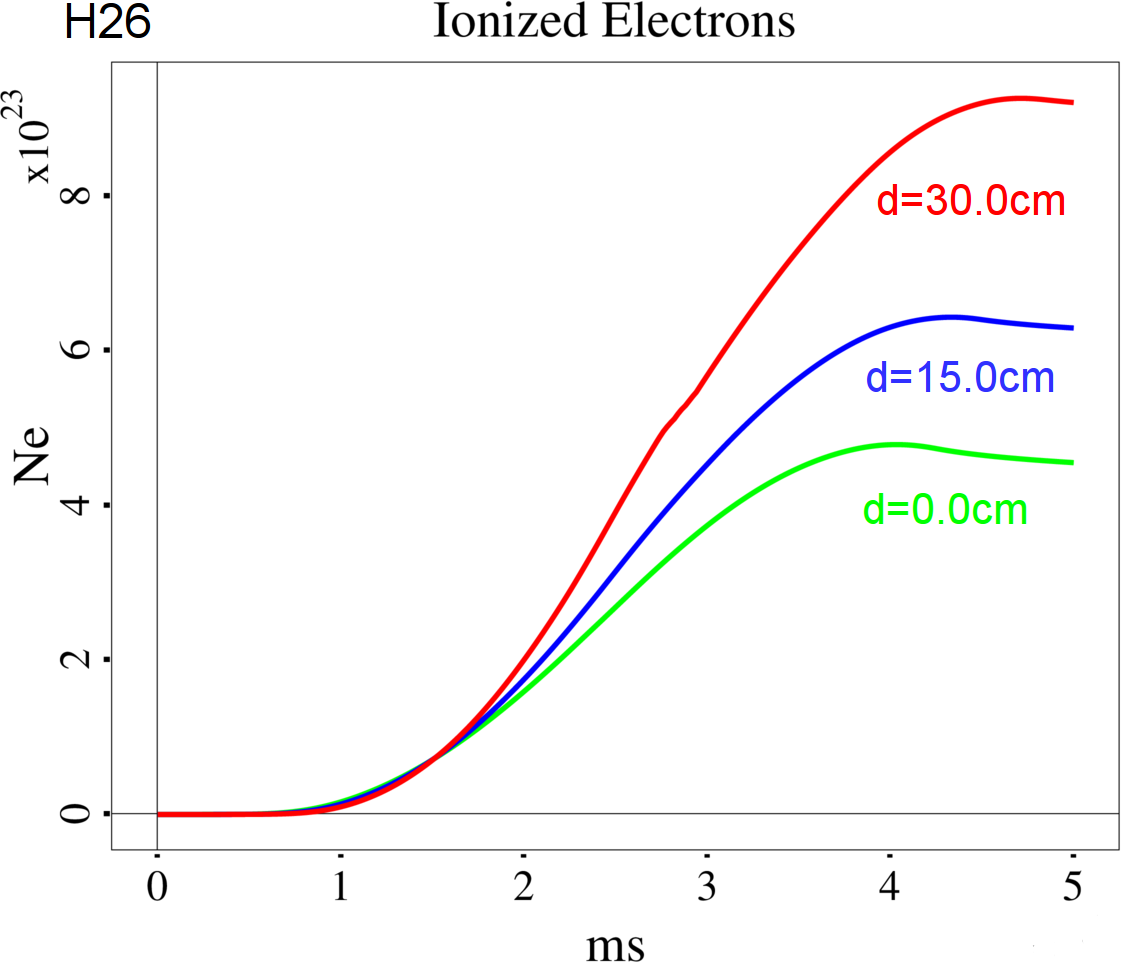}
		\caption{H26 L-mode}
		\label{fig:H26l_noff}
	\end{subfigure}
	\hspace{0.00cm}
	\begin{subfigure}[b]{0.50\textwidth}
		\includegraphics[width=1.00\textwidth]{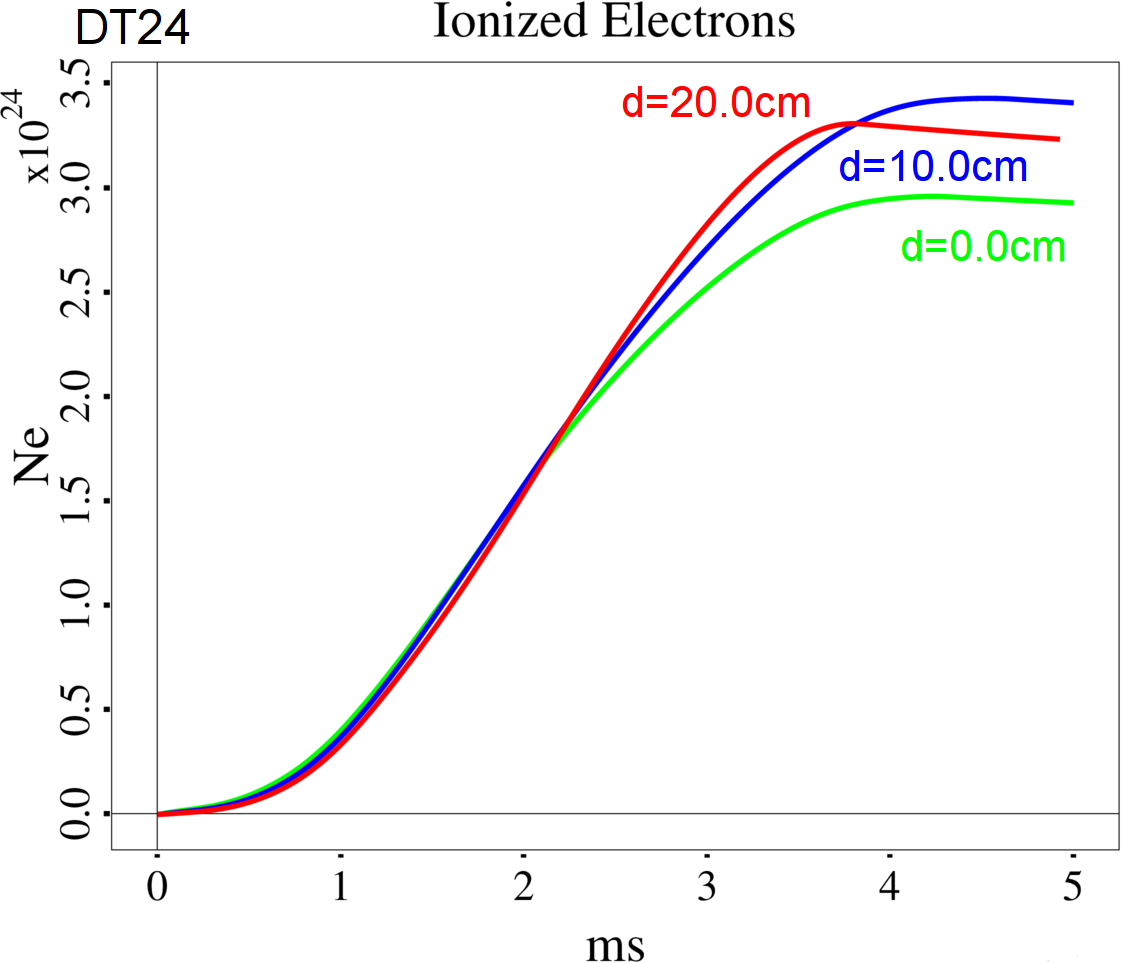}
		\caption{DT24}
		\label{fig:DT24_noff}
	\end{subfigure}
	}
	\caption{Comparison of thermal, radiated and Ohmic energies (\ref{fig:H26l_Toff} and \ref{fig:DT24_Toff}) for
	deposition offsets for H26 L-mode and DT24 (low and high thermal energies) shows minor impact on thermal quench.
	The ionized electron count (\ref{fig:H26l_noff} and \ref{fig:DT24_noff}) reflects the increasing ablation with
	deposition offset.  The seemingly greater impact on the DT24 results (\ref{fig:DT24_Toff} and
	\ref{fig:DT24_noff}) reflect the impact of an increasing assimilation fraction approaching 1 (with complete burn through for
	\clrr{d=20.0cm} at t=3.86ms.) rather than the deposition offset.}
	\label{fig:nToff}
\end{figure}

For the deposition offset, we return to the r$_f$=2.5mm uniform pencil beam plume of 40.0cm length.

Figures~\ref{fig:nToff} show a comparison of the thermal, radiated, and Ohmic energies, and a comparison of the number
of ionized Electrons.  Again, we see that for increasing evaluation offset, the thermal energy evolution does not change
significantly.  The differences again can be attributed primarily to the greater burn through of the plasma plume with
increasing offset, thereby delaying the quench.  It can be inferred from the comparison of the ionized electrons, which
results dominantly from the deuterium (95\% of pellet), that more ablation occurs for increasing offset.  For
DT24(\ref{fig:DT24_Toff} and \ref{fig:DT24_noff}) show complete burn through for \clrr{d$_{offset}$=20.0cm}.  

Tables~\ref{tab:depo} thermal quench times use the thermal quench time for the no offset \clrlg{0.0cm} case to
measure the assimilation and radiation fractions.  This table clearly shows the increasing ablation with increasing
deposition offset.  However, the radiation fraction and q=2 quench times are only modestly effected.

\section*{Acknowledgement}
This material is based upon work supported by ITER Contract \#IO/IA/20/4300002130.

\bibliographystyle{plain}
\bibliography{ref}

\end{document}